\documentclass[12pt]{article}

\usepackage{graphics,graphicx,cite,amssymb,epsfig,float,psfrag}
\usepackage[usenames]{color}
\usepackage{rotating}
\newcommand{\lsim}{\raisebox{-0.13cm}{~\shortstack{$<$ \\[-0.07cm] $\sim$}}~} 
\newcommand{\gsim}{\raisebox{-0.13cm}{~\shortstack{$>$ \\[-0.07cm] $\sim$}}~} 
\newcommand{\ra}{\rightarrow} 
 
\newcommand{\tb}{\tan\beta} 
\newcommand{\beq}{\begin{eqnarray}} 
\newcommand{\eeq}{\end{eqnarray}} 
\newcommand{\s}{\\ \vspace*{-4mm}}

\newcommand{\non}{\nonumber}

\begin{document}
\thispagestyle{empty}

\vspace{1cm}

\hfill CERN--PH--TH/2012--185

\hfill LPT--ORSAY--12/63

\vspace*{1.5cm}

\begin{center}

\mbox{\large\bf The Higgs sector of the phenomenological MSSM} \\

\vspace*{.2cm}

\mbox{\large\bf in the light of the Higgs boson discovery}

\vspace*{.8cm}

A.~Arbey$^{a,b,c}$, M.~Battaglia$^{c,d,e}$, A.~Djouadi$^{c,f}$ and  F.~Mahmoudi$^{c,g}$ 

\vspace*{8mm}

{\small 
    
$^a$ Universit\'e de Lyon, France; Universit\'e Lyon 1, CNRS/IN2P3, UMR5822 IPNL, 
F-69622~Villeurbanne Cedex, France.  

$^b$ Centre de Recherche Astrophysique de Lyon, Observatoire de Lyon, Saint-Genis Laval Cedex, 
F-69561, France; CNRS, UMR 5574; Ecole Normale Sup\'erieure de Lyon, Lyon, France. 

$^c$ CERN, CH-1211 Geneva 23, Switzerland. 

$^d$ Santa Cruz Institute of Particle Physics, University of California, Santa Cruz,
CA 95064, USA. 

$^e$ Lawrence Berkeley National Laboratory, Berkeley, CA 94720, USA. 

$^f$ Laboratoire de Physique Th\'eorique, Universit\'e Paris XI and CNRS,
F--91405 Orsay, France.

$^g$ Clermont Universit\'e, Universit\'e Blaise Pascal, CNRS/IN2P3,\\LPC, BP 10448, 63000 Clermont-Ferrand, France. 

}
\end{center}

\vspace{.4cm}

\begin{abstract}
  	
The long awaited discovery of a new light scalar boson at the LHC opens up a new
era of studies of the Higgs sector in the Standard Model  and  in its extensions.
In this paper we discuss the consequences of the observation of a light Higgs
boson with the mass and rates reported by the ATLAS and CMS collaborations on
the parameter space of the phenomenological MSSM, including also the LHC searches 
for heavier Higgs bosons and supersymmetric
particle partners,  as well as the constraints from $B$--physics and dark matter. 
We explore the various regimes of the MSSM Higgs sector depending  on the parameters 
$M_A$ and $\tan \beta$ and show that only two of them are still allowed by all 
present experimental constraints:  the decoupling regime in which there is only one 
light and standard--like Higgs boson while the heavier  Higgs states decouple from 
gauge bosons, and the supersymmetric regime, in which there are light supersymmetric 
particle partners which might affect the decay properties of the light Higgs boson, 
in particular its di-photon and invisible decays. 

\end{abstract} 

\thispagestyle{empty}
\vspace*{1cm}

\newpage

\section{Introduction}

Results of the search for the Higgs bosons at the LHC with the 10~fb$^{-1}$ data
collected in 2011 at 7~TeV and 2012 at 8~TeV have just been presented  by the
ATLAS\cite{ATLAS:2012zz} and CMS~\cite{CMS:2012zz} collaborations and there is
now a  5$\sigma$ evidence by each of the experiments for a new particle with a mass 
of $\approx 126$~GeV. Complementary evidence  is also provided by the updated
combination of the Higgs searches performed by the CDF and D0 collaborations at
the  Tevatron~\cite{Tevatron:2012zz}, which has also been just released. As we
are entering an era of Higgs studies, these results have deep implications for
the Minimal Supersymmetric Standard Model (MSSM).
The implications of a Higgs boson with a  mass value around  $126$ GeV in the context 
of Supersymmetry have been already  widely 
discussed~\cite{ABDMQ,Arbey:2011aa,Carena:2011aa,papers-2011} since the first
evidence of a signal at the LHC was presented at the end of 2011. In particular, 
we have discussed the consequences of the  value of $M_h$ for the unconstrained 
phenomenological MSSM (pMSSM) with 22 free parameters~\cite{pMSSM}, for
constrained MSSM scenarios such as the minimal  gravity, gauge and anomaly
mediated SUSY--breaking  models, and in scenarios in which  the supersymmetric 
spectrum is extremely heavy~\cite{ABDMQ}.  We have shown that only when the
SUSY--breaking scale is very large or the  mixing in the stop sector is
significant the observed $M_h$ value can be accommodated in these models. This
disfavours many  constrained scenarios such as the minimal anomaly and gauge
mediated SUSY--breaking models and some (even more constrained)  versions of the
minimal super-gravity model. 

In this paper, we extend the previous study in new directions afforded by the
improved data from the LHC experiments. First, we refine our analysis of the
implications of the value $M_h \approx 126$ GeV for the decoupling regime by
considering different types of stop mixing scenarios which significantly  affect
the maximal mass value and we explore the implications of a broader range of the 
top quark mass value, $m_t$ = (173$\pm$3)~GeV, on $M_h$.  Then, we analyse in detail 
the implications of the ATLAS and CMS searches for the heavier MSSM Higgs bosons, 
the CP--even $H$, a pseudoscalar $A$ and two charged $H^\pm$ states. 
In particular,  we discuss the  $A/H/h \to \tau^+\tau^-$ for 
the neutral  and the $t \to bH^+ \to b \tau \nu$ searches for the charged states  to 
further constrain the $[M_A,\tb]$ parameter space, where $\tb$ is the ratio of 
the vacuum expectation values of the two Higgs doublet fields. We also discuss the 
effect of the recent LHCb results for the decay $B_s \to \mu^+\mu^-$  on the Higgs 
sector, as well as the super-particle LHC searches and the dark matter constraints.  
Most importantly, we study other regimes than the decoupling regime of the pMSSM: the
anti-decoupling regime for low $M_A$ in  which  the roles of the $h$ and $H$
bosons are reversed,  the intense coupling regime in which the three neutral
particles  $h,A,H$ are rather close in mass, the intermediate regime at 
relatively low $\tb$  in which the couplings of $H,A$ to gauge+Higgs bosons are
not too suppressed and the vanishing coupling regime in which  the coupling of
the $h$ state to   bottom quarks or gauge bosons is suppressed.  Using the
latest ATLAS, CMS and Tevatron data, we show that all these scenarios are now
almost ruled out.  Finally, we start studying the implications of the rates
reported by the LHC experiments in the $\gamma \gamma$ and $ZZ$ final states 
used to obtain the Higgs boson signal and we comment on the $b \bar b$ final
state to which the Tevatron is most sensitive. We perform a full scan of the
pMSSM parameter space in order to delineate the regions which fit best the
experimental data, including a possible enhancement of the $h\to \gamma\gamma$
rate. 

The paper is organised as follows. First, we briefly describe the pMSSM and its  
Higgs sector with its various regimes and summarise the Higgs decays and the 
production cross sections at the LHC. In section 3,  we present the analysis of 
these different Higgs regimes and the implications on the pMSSM parameters in the 
light of the LHC Higgs discovery and constraints. Section 4 has a short conclusion.

\section{The theoretical set-up}

\subsection{The pMSSM Higgs sector}

In the MSSM the Higgs sector is extended to contain five Higgs particles 
\footnote{ For a review of the MSSM Higgs sector~\cite{Review2}. 
For reviews on the radiative corrections in the MSSM Higgs sector and a complete 
set of references, see~\cite{rad-cor}}. 
The lightest $h$ boson has in general the properties of the Standard Model (SM)
Higgs boson \footnote{For a review of the SM Higgs boson, see~\cite{Review1}} 
and is expected to have a mass $M_h \lsim 115$--135~GeV depending on the MSSM 
parameters, in particular, the ratio $\tan\beta$ of the vacuum expectation values  
of the two Higgs doublet fields that break the electro-weak symmetry in the MSSM.  

By virtue of supersymmetry,  only two parameters are needed to describe the
Higgs sector at  tree--level. These can be conveniently chosen to be the
pseudoscalar boson mass $M_A$ and the ratio of vacuum expectation values of the
two Higgs fields that break the symmetry, $\tan\beta= v_2/v_1$. However, 
accounting for the radiative corrections to the Higgs sector, known to play an
extremely important role~\cite{rad-cor},  all soft SUSY--breaking parameters 
which are of ${\cal O}(100)$  in addition to those of the SM,  become relevant. 
This makes any phenomenological analysis in the most general MSSM a very
complicated  task. A phenomenologically more viable MSSM framework, the 
pMSSM, is defined by adopting the following assumptions:  $i)$ all
soft SUSY--breaking parameters are real and there is no new source of
CP--violation; $ii)$ the matrices for the sfermion masses and for the trilinear
couplings are all diagonal,  implying no flavor change   at  tree--level; and
$ii)$ the soft SUSY--breaking masses and trilinear couplings of the first and
second sfermion generations are the same  at the electro-weak symmetry breaking
scale. Making these three assumptions will lead to only 22 input parameters in
the pMSSM: 

\begin{itemize} 
\vspace*{-3mm}
\item[--] $\tan \beta$: the ratio of the vevs is expected to lie in the range $1 \lsim \tb \lsim m_t/m_b$;\vspace*{-3mm} 

\item[--] $M_A$: the pseudoscalar Higgs mass that ranges from $M_Z$ to the
SUSY--breaking scale;\vspace*{-3mm}

\item[--] $\mu$: the Higgs--higgsino (supersymmetric) mass parameter (with both
signs);\vspace*{-3mm} 

\item[--] $M_1, M_2, M_3$: the bino, wino and gluino mass parameters;\vspace*{-3mm}

\item[--] $m_{\tilde{q}}, m_{\tilde{u}_R}, m_{\tilde{d}_R},  m_{\tilde{l}},
 m_{\tilde{e}_R}$: the first/second generation sfermion mass
 parameters;\vspace*{-3mm}

\item[--]  $A_u, A_d, A_e$: the first/second generation trilinear
couplings;\vspace*{-3mm}

\item[--] $m_{\tilde{Q}}, m_{\tilde{t}_R}, m_{\tilde{b}_R},  m_{\tilde{L}},
 m_{\tilde{\tau}_R}$: the third generation sfermion mass parameters;\vspace*{-3mm}

\item[--] $A_t, A_b, A_\tau$: the third generation trilinear
couplings.\vspace*{-3mm}
\end{itemize} 

Such a model has more predictability and it offers an adequate framework 
for phenomenological studies. In general,  only a small subset of the parameters 
appears when looking at a given sector of the pMSSM, such as the Higgs sector 
in this case. 
Some of these parameters will enter the radiative corrections to the Higgs boson
masses and couplings.  At the one--loop level, the $h$ boson mass receives
corrections that  grow as the fourth power of the top quark mass $m_t$ (we use
the running $\overline{\rm MS}$ mass to re-sum  some higher order  
corrections)  and logarithmically with the SUSY--breaking scale   or common
squark mass $M_S$; the trilinear coupling in the stop sector $A_t$
plays also an important role. The leading part of these corrections reads~\cite{RC-1loop}
\beq 
\epsilon = \frac{3\, \bar{m}_t^4}{2\pi^2 v^2\sin^ 2\beta} \left[ \log
\frac{M_S^2}{\bar{m}_t^2} + \frac{X_t^2}{2\,M_S^2} \left( 1 -
\frac{X_t^2}{6\,M_S^2} \right) \right]. \label{epsilon} 
\eeq 
We  have defined  the SUSY--breaking scale $M_S$ to be the geometric average 
of the two stop masses (that we take $\lsim 3$ TeV not to introduce excessive 
fine-tuning)
\beq 
M_S = \sqrt{ m_{\tilde t_1} m_{\tilde t_2}} 
\eeq  
and introduced the  mixing parameter $X_t$ in the stop sector (that we assume $\lsim 
3M_S$),  
\beq 
X_t = A_t -\mu \cot\beta .
\eeq
The radiative corrections have a much larger impact and maximise the
$h$ boson mass in the so--called ``maximal mixing" scenario, where the 
trilinear stop coupling in the $\overline{\rm DR}$ scheme is  
\beq 
{\rm maximal~mixing~scenario~:} \quad X_t =
\sqrt{6}\, M_S .
\eeq
In turn,  the radiative corrections are much smaller for small values of 
$X_t$,  i.e. in the 
\beq
{\rm no~mixing~scenario:}\quad X_t=0 .
\eeq
An intermediate scenario is when $X_t$ is of the same order as $M_S$ 
which is sometimes called the 
\beq
{\rm typical~mixing~scenario:}\quad X_t=M_S .
\eeq
These mixing scenarios have been very often used as benchmarks for the analysis 
of MSSM Higgs phenomenology \cite{benchmarks}. The maximal mixing scenario has
been particularly privileged since it gives a reasonable estimate  of the upper
bound on the $h$ boson mass,  $M_h^{\rm max}$. We will  discuss these scenarios
but, compared to the work of Ref.~\cite{benchmarks}, we choose here to vary the
scale $M_S$. Together with the requirements on $X_t$ in eqs.~(4--6),  we adopt
the following values  for the  parameters entering the pMSSM Higgs sector, 
\begin{eqnarray}
A_t=A_b \,, \ M_2  \simeq 2 \,M_1 = |\mu|=\frac15 M_S \, , \ M_3=0.8 \,M_S\, ,
\label{pbenchmark}
\end{eqnarray}
and vary the basic inputs $\tb$ and $M_A$. For the values $\tb=60$ and
$M_A=M_S=3$ TeV and a top quark pole of mass of $m_t=173$ GeV, we would obtain a 
maximal Higgs mass value  $M_h^{\rm max} \approx 135$ GeV for maximal mixing once 
the full set of known radiative corrections up to two loops is implemented \cite{adkps}. 
In the no--mixing  and  typical mixing scenarios, one obtains  much smaller values, $M_h^{\rm max} 
\approx  120$ GeV and $M_h^{\rm max} \approx 125$ GeV, respectively.  Scanning
over the soft SUSY--breaking parameters, one may  increase these $M_h^{\rm max}$
values by up to a few  GeV. 

It is important to note that the dominant two--loop corrections have been
calculated   in the $\overline{\rm DR}$ scheme \cite{BDSZ} and implemented in
the codes  {\tt Suspect}~\cite{suspect} and {\tt SOFTSUSY}~\cite{Allanach:2001kg}
that we will use here  for the MSSM spectrum, but also  in the on--shell scheme
\cite{HHW} as  implemented in  {\tt FeynHiggs} \cite{Heinemeyer:1998yj}.  In
general, the results for $M_h$ in the two scheme differ  by at most 2 GeV, which
we take as a measure of the missing higher order effects.   Quite recently,  the
dominant three--loop contribution to $M_h$ has been calculated and found to be
below 1~GeV \cite{3loop}. Thus, the mass of the lightest  $h$ boson can be
predicted with an accuracy of $\Delta M_h \sim 3$~GeV and this is  the
theoretical uncertainty on $M_h$ that we assume.

\subsection{The various regimes of the pMSSM}

The spectrum in the various regimes of the pMSSM Higgs sector~\cite{Review2}, 
depends on the values of $M_A$ and also on $\tb$, and that we will confront 
to the latest LHC and Tevatron data in this paper. 

We start from the \underline{decoupling regime} \cite{Decoupling} that has
been  already mentioned and which in principle occurs for large values of $M_A$ but
is reached  in practice at $M_A \gsim 300$~GeV for low $\tb$ values and
already  at $M_A \gsim M_h^{\rm max}$ for $\tb \gsim 10$. In this case,  the
CP--even $h$ boson reaches its maximal mass value $M_h^{\rm max}$ and its
couplings to fermions and gauge bosons (as well as its self--coupling) become
SM--like. The heavier $H$ boson has approximately the same mass as the $A$ boson
and its interactions are similar, i.e. its couplings to gauge bosons almost
vanish and the couplings to bottom (top) quarks and $\tau$ leptons   fermions
are (inversely) proportional to $\tb$.  Hence, one will have a SM--like Higgs
boson $h \equiv H_{\rm SM}$  and two pseudo-scalar (like) Higgs particles,
$\Phi=H,A$.  The $H^\pm$  boson is also degenerate in mass with the $A$ boson
and the intensity of its couplings to fermions is similar.  Hence, in the
decoupling limit, the heavier $H/A/H^\pm$  bosons  almost decouple and the MSSM
Higgs sector reduces effectively to the SM Higgs sector, but with a light $h$
boson. \s

The \underline{anti--decoupling regime} \cite{Antidecoup} occurs for a  light
pseudo-scalar Higgs boson, $M_A \lsim M_h^{\rm max}$, and  is exactly opposite to
the decoupling regime. The roles of the $h$ and $H$ bosons are reversed and at
large $\tb$ values, the $h$ boson is degenerate in mass with the pseudo-scalar
$A$, $M_h \simeq M_A$, while the $H$ boson has a mass close to its minimum which
is in fact $M_h^{\rm max}$. Because of the upper bound on $M_h$, all Higgs
particles are light. Here,  it is the $h$ boson which has  couplings close to
those of $A$, $\Phi=h,A$,   while the $H$ boson couplings are SM--like, $H\equiv
H_{\rm SM}$.\s

The \underline{intense--coupling regime} \cite{intense}  occurs when the
mass of the pseudo-scalar $A$ boson is close to $M_h^{\rm max}$. In this case,
the three neutral Higgs bosons $h,H$ and $A$ (as well as the charged Higgs
particles) have comparable masses, $M_h \sim M_H \sim M_A \sim M_h^{\rm max}$.
The mass degeneracy is more effective when $\tb$ is large. Here,  both the $h$
and $H$ bosons have still enhanced couplings to $b$--quarks and $\tau$ leptons 
and suppressed couplings to gauge bosons and top quarks, as is the  pseudo-scalar
$A$. Hence, one approximately has three pseudo-scalar like Higgs particles, $\Phi
\equiv h,H,A$ with mass differences of the order of 10--20 GeV.\s

The \underline{intermediate--coupling regime} occurs for low values of $\tb$,
$\tb \lsim 5$--10,  and a not too heavy pseudo-scalar Higgs boson, $M_A \lsim
300$--500 GeV \cite{Review2}. Hence, we are not yet in the 
decoupling regime and  both CP--even Higgs bosons have non--zero couplings to 
gauge bosons and their couplings to down--type (up--type) fermions (as is the case 
for the  pseudoscalar $A$ boson) are not strongly enhanced (suppressed) since $\tb$ 
is not too large. This scenario is already challenged by LEP2 data which call for  
moderately large values of $\tb$. \s

The \underline{vanishing--coupling regime}  occurs for relatively large
values of $\tb$ and intermediate to large $M_A$ values, as well as for specific
values of the other MSSM parameters. The latter parameters, when  entering the
radiative corrections, could lead to a strong suppression of the couplings of
one of the CP--even Higgs bosons to fermions or gauge bosons, as a result of the
cancellation between tree--level terms and radiative corrections
\cite{vanishing}. An example of such a situation is the small $\alpha_{\rm eff}$
scenario which has been used as a benchmark \cite{benchmarks} and in which the
Higgs to $b\bar b$ coupling is strongly suppressed.\s 

\begin{figure}[!h]
\begin{center}
\includegraphics[width=10.cm]{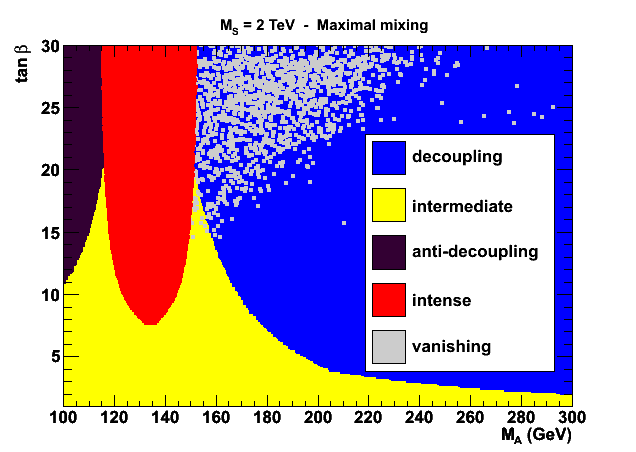}
\end{center}
\vspace*{-5mm}
\caption[]{\small The parameter space for the various regimes of the MSSM Higgs
sector as defined in the text and in eq.~(8) in the $[M_A,\tb]$ plane; the  
maximal  mixing scenario with $M_S=2$ TeV is adopted.}
\label{fig:regimes}
\end{figure}

Within the plane $[M_A,\tan\beta]$, the parameter space in which the above
regimes of the pMSSM Higgs sector occur are displayed in Figure~\ref{fig:regimes}. 
We have chosen the usual maximal mixing scenario with $M_S=2$ TeV and the other SUSY
parameters as in eq.~(\ref{pbenchmark}), except for the vanishing coupling
scenario, where we have scanned over the SUSY parameters, and only $\approx 5 \times 10^{-4}$ 
of the scanned points fulfil its requirements. The following conditions have been imposed: 
\beq
{\rm decoupling~regime} &:& \cos^2(\beta-\alpha) \leq 0.05 \non \\
{\rm anti-decoupling~regime} &:& \cos^2(\beta-\alpha) \geq 0.95 \non \\
{\rm intermediate-coupling~regime} &:&  0.05 \leq \cos^2(\beta-\alpha) \leq 
0.7, ~~\tan\beta \leq 10 \non \\
{\rm intense-coupling~regime} &:& M_A\lsim 140~{\rm GeV},~ g_{hbb}^2~{\rm and}~g_{Hbb}^2 \geq 50 \non \\
{\rm vanishing-coupling~regime} &:&M_A \gsim 200~{\rm GeV},~
g_{hbb}^2~{\rm or}~g_{hVV}^2 \leq 0.05 .
\eeq

In addition, we have to consider the \underline{SUSY regime}, in which some 
SUSY particles such as the charginos, neutralinos as well as the third generation
sleptons  and squarks, could be light enough to significantly affect the
phenomenology of the MSSM Higgs bosons. For instance, light sparticles could
substantially contribute to the loop induced production and decays modes of the
lighter $h$ boson~\cite{ST-gg,ST-pp}  and could even appear (in  the case of
the lightest neutralino) in its decay product as will be discussed below.  

\subsection{ Higgs decays and production in the pMSSM}

For the relatively large values of $\tb$ presently probed at the LHC, 
$\tb \gsim 7$ as discussed below, the couplings of the non--SM like Higgs
bosons to $b$ quarks and $\tau$ leptons are so strongly enhanced and those to
top quarks and gauge bosons suppressed, that the pattern becomes as  simple as
the following (more details can be found in Ref.~\cite{Review2}):

-- The $\Phi=A$ or $H/h$ bosons  in the decoupling/anti-decoupling limit  decay 
almost exclusively into $b\bar b$ and $\tau^+\tau^-$ pairs,  with branching
ratios of, respectively, $\approx 90\%$ and $ \approx 10\%$, and all  other
channels are suppressed to a level where their branching ratios are negligible. 

-- The $H^\pm$ particles decay into fermion pairs: mainly $ H^+  \to t \bar{b}$
and $H^+ \to \tau \nu_{\tau}$ final states for $H^\pm$ masses, respectively,
above and below the $tb$ threshold. 

-- The CP--even  $h$ or $H$ boson, depending on whether we are in the decoupling
or anti-decoupling regime, will have the same decays as the SM Higgs boson.  For
$M_{h/H}\approx 126$ GeV, the main decay mode will be the  $b\bar{b}$ channel
with a $\sim$ 60\% probability, followed by the decays into $c\bar{c}$,
$\tau^+\tau^-$ and the loop induced decay into gluons  with $\sim$ 5\% branching
ratios. The $WW^*$ decay  reaches the level of 20\%, while the rate for $ZZ^*$
is  a few times $10^{-2}$. The important loop induced $\gamma \gamma$ decay
mode  which leads to  clear signals at the LHC have rates of ${\cal
O}(10^{-3})$.

 In the intense--coupling regime,  the couplings of both $h$ and  $H$  to gauge
bosons and up--type fermions are suppressed and those  to down--type fermions
are enhanced. The branching ratios of the $h$ and $H$ bosons to $b\bar{b}$ and
$\tau^+\tau^-$ final states are thus the dominant ones, with values as in the
case of the pseudoscalar $A$ boson.   In the intermediate--coupling regime,  
interesting decays of $H,A$ and $H^\pm$ into gauge and/or Higgs bosons occur, 
as well as $A/H \to t\bar t$ decays, but they are suppressed in general.
Finally, for the rare vanishing--coupling regime when the Higgs couplings to
$b$--quarks and eventually $\tau$--leptons accidentally vanish, the outcome is
spectacular for the $h$ boson:  the $WW^*$ mode becomes dominant and  followed
by $h \to gg$, while the interesting $h \to \gamma \gamma$ and $h\to ZZ^*$ decay
modes  are enhanced. 

In the case of the SM--like Higgs particle (that we assume now to be the $h$
boson),  there are two interesting  scenarios which might make its
decays rather different.  
First we have the scenario with the Higgs bosons decaying into supersymmetric 
particles. Because most sparticles must be heavier than about 100 GeV, there is
no SUSY decays of the $h$ boson except for  the invisible channel into a pair of
the lightest neutralinos, $h \to \chi_1^0\chi_1^0$. This is particularly true
when the gaugino mass universality relation  $M_2 \sim 2M_1$  is relaxed,
leading to light $\chi_1^0$ states  while the LEP2 bound, $m_{\chi_1^\pm} \gsim
100$ GeV,  still holds.  In the decoupling limit,  the branching ratio of the 
invisible  decay can reach the level of a few 10\%. Decays of the heavier
$A/H/H^\pm$ bosons, in particular into charginos, neutralinos, sleptons and top
squarks, are in turn possible. However, for  $\tb \gsim 10$, they are strongly
suppressed.

The second scenario of interest occurs when SUSY particles contribute to 
loop-induced Higgs decays.  If scalar quarks  are relatively 
light, they can lead to sizable contributions to the  decays $h \to gg$ and 
$h\to \gamma \gamma$.  Since scalar quarks have Higgs couplings  that are not 
proportional to their masses, their contributions are damped by loop factors $1/m_{\tilde
Q}^2$ and decouple from the vertices contrary to SM quarks. Only when $m_{\tilde
Q}$ is not too large compared to $M_{h}$ that  the contributions are 
significant~\cite{ST-gg}.  This is particularly true for the $\tilde t_1$
contributions  to  $h \to gg$,  the reasons being that large $X_t$ mixing leads
to a  $\tilde{t}_1$ that is much  lighter than all  other  squarks  and that the
$h$ coupling to  stops  involves a component which is proportional to $m_t X_t$
and, for large  $X_t$,  it can be strongly enhanced. Sbottom mixing, $ \propto
m_b X_b$, can also be sizable for large $\tb$ and $\mu$ values and can lead to
light $\tilde b_1$ states with strong couplings to the $h$ boson. In $h \to
\gamma \gamma$ decay, there are in addition  slepton loops,   in particular
$\tilde \tau$ states which behave like scalar bottom quarks and have a strong mixing 
at high $\mu\tb$, can make a large impact on the  decay rate. Besides, chargino loops
also enter the $h \to \gamma \gamma$ decay mode but their contribution is in
general smaller  since the Higgs--$\chi \chi$ couplings cannot be  strongly
enhanced. 

For the evaluation of the decay branching ratios of the MSSM Higgs bosons, we
use the program {\tt HDECAY}~\cite{Djouadi:1997yw}, which incorporates all decay channels
including those involving super-particles and the most important sets of higher
order corrections and effects. 

Coming to Higgs boson production at the LHC, for a SM--like particle $H_{\rm SM}$
there are essentially four mechanisms for single  production \cite{Review1}.
These are  $gg$ fusion,  $gg \to H_{\rm SM}$,  vector boson fusion, $qq \to
H_{\rm SM}qq$,  Higgs-strahlung,  $q\bar q \to H_{\rm SM}V$ and $t\bar t$
associated Higgs production,  $pp \to t\bar tH_{\rm SM}$.  The $gg \to H_{\rm
SM}$ process  proceeds  mainly through a heavy top quark loop and  is by far the
dominant  production mechanism at the LHC. For a  Higgs boson with a  mass of
$\approx 126$ GeV,  the cross section is more than one order of magnitude larger
than in  the other processes. Again for $M_{H_{\rm SM}} \approx 126$ GeV, the
most efficient  detection channels  are  the clean but rare $H \to \gamma
\gamma$ final states, the modes $H\to ZZ^* \to 4\ell^\pm$, $H\to WW^{(*)}\to
\ell \ell \nu \nu$ with $\ell=e,\mu$ and, to a lesser extent, also $H_{\rm SM}
\to \tau^+\tau^-$. At the LHC and, most importantly, at the Tevatron one is
also sensitive to $q\bar q \to H_{\rm SM}+W/Z \to  b\bar
b+W/Z$  with $W\to \ell \nu$ and $Z\to \ell \ell, \nu \bar \nu$. 

For the MSSM Higgs bosons, the above situation holds for the $h(H)$ state in the
\mbox{(anti-)} decoupling regime.  Since $AVV$ couplings are absent, the $A$
boson cannot be produced in Higgs-strahlung and vector boson fusion and the
rate  for $pp \to t\bar t A$ is strongly suppressed.  For the $\Phi=A$ and
$h(H)$ states, when we are in the  (anti-)decoupling limit,  the $b$ quark will
play an important role for large $\tb$ values as the $\Phi bb$  couplings are
enhanced.  One then has to take into account the $b$--loop contribution  in the
$gg \to \Phi$  processes which becomes the dominant component in the MSSM and
consider  associated Higgs production with $b\bar{b}$ final states, $pp \to b
\bar b + \Phi$  which become the dominant channel in the MSSM. The latter
process is in fact equivalent to $b\bar b \to \Phi$ where the $b$--quarks are
taken from the proton in a five active flavor scheme. As the $\Phi$ bosons decay
mainly into $b\bar b$ and $\tau^+\tau^-$ pairs, with the former being swamped by
the QCD  background, the most efficient detection channel would be $pp \to \Phi
\to \tau^+ \tau^-$. This process receives contributions from both the $gg \to 
\Phi$ and $b\bar b \to \Phi$ channels. 

These processes also dominate the $h/H/A$ production in the intense coupling
regime. In fact, in the three regimes above, when all  processes leading to
$\tau^+\tau^-$ final states are added up, the rate is $2\times
\sigma(gg+b\bar b \to A)\times{\rm BR}(A\to \tau^+\tau^-)$. In the intermediate
coupling regime, these process have very low cross sections as for 3--$5 \leq
\tb \leq 7$--10,  the $\Phi bb$ couplings  are not enough enhanced and the
$\Phi tt$ ones that  control the  $gg$ fusion rate are still suppressed. 

Finally, for the charged Higgs boson, the dominant channel is the production
from top quark decays, $t \to H^+ b$, for masses not too close to
$M_{H^\pm}=m_t\!-\! m_b$. This is true in particular at low or large $\tb$ values 
when the $t\to H^+b$ branching ratio is significant. 

The previous discussion on MSSM Higgs production and detection at the LHC might
be significantly altered  if scalar quarks, in particular $\tilde t$ and $\tilde b$, 
are light enough. Indeed, the $Hgg$ and $hgg$ vertices in the MSSM are mediated
not only by the $t/b$ loops but also by loops involving their partners similarly
to the Higgs photonic decays.  The $gg\to h$ cross section in the decoupling
regime  can be  significantly altered by light stops and a strong mixing
$X_t$ which enhances the $h\tilde t_1 \tilde t_1$ coupling. The cross section
times branching ratio $\sigma( gg \ra h) \times {\rm BR}(h \ra \gamma \gamma)$
for the lighter $h$ boson at the LHC could be thus  different from the SM case,
even in the decoupling limit in which the $h$ boson is supposed to be
SM--like~\cite{ST-gg}. 

Finally, we should note that  in the scenario in  which the Higgs bosons, and in
particular the lightest one $h$, decay into invisible lightest neutralinos,  $h
\to \chi_1^0 \chi_1^0$,  the observation of the final state  will be 
challenging but possible at the LHC with a higher energy and more statistics.
This scenario  has recently been discussed in detail in
Refs.~\cite{AlbornozVasquez:2011aa,Arbey:2011aa}.

\section{Analysis and results} 

\subsection{pMSSM scans and software tools}

The analysis is based on scans of the multi-parameter MSSM phase space. The input 
values of the electro-weak parameters, i.e.\ the top quark pole mass, the 
$\overline{\rm MS}$ bottom quark mass, the  electro-weak gauge boson masses, 
electromagnetic and strong coupling constants defined at the scale $M_Z$, 
are given below with  their 1$\sigma$ allowed ranges \cite{PDG},
\begin{eqnarray}
m_t=(173 \pm 1)~{\rm GeV}, \ \bar m_b (\bar m_b)=(4.19^{+0.18}_{-0.06})~{\rm GeV}, \non \\
M_Z=(91.19 \pm 0.002)~{\rm GeV}, \ M_W=(80.42 \pm 0.003)~{\rm GeV},  \nonumber \\ 
\alpha (M_Z^2) =1/127.916 \pm 0.015, \ \alpha_s(M_Z^2) =0.1184 \pm 0.0014\, . 
\label{SM-inputs} 
\end{eqnarray}
The pMSSM parameters are varied in an uncorrelated way in flat scans, within the
following ranges:
\begin{eqnarray}
1\leq \tb \leq 60 \, , ~~~~~~~~~~~ \,  \nonumber \\
  50~{\rm GeV} \leq M_A \leq 3~{\rm TeV}\, , ~~~~~~~~ \nonumber \\ 
\ -10~{\rm TeV} \leq A_f \leq 10 ~{\rm TeV} \, , ~~~~~~    \nonumber \\
~~~50~{\rm GeV} \leq m_{\tilde f_L}, m_{\tilde f_R}, M_3 
\leq 3.5~{\rm TeV}\, , \nonumber \\  
50~{\rm GeV} \leq M_1, M_2, |\mu| \leq 2.5~{\rm TeV}
\label{scan-range}
\end{eqnarray}
to generate a total of $6\times 10^7$  pMSSM points. The scan range is
explicitly  chosen to include the various mixing scenarios in the Higgs section
discussed in section 2.1:  the  maximal mixing, no--mixing and typical mixing
scenarios. Additional $10^7$ points are generated  in specialised scans used for
the studies discussed later in section~3.5. We select  the set of points 
fulfilling constraints from flavour physics and lower energy searches  at LEP2
and the Tevatron, as discussed in~Ref.~\cite{Arbey:2011un}, to which we refer
also for details on the scans.  
We highlight here the tools most relevant to this study. The SUSY mass spectra
are generated with {\tt SuSpect}~\cite{suspect} and  {\tt SOFTSUSY
3.2.3}~\cite{Allanach:2001kg}.    The superparticle partial decay widths and 
branching  fractions are computed using the program {\tt SDECAY
1.3}~\cite{sdecay}. The flavour observables  and dark matter relic density are
calculated with {\tt SuperIso Relic v3.2}~\cite{flavor}. 

The Higgs production cross sections at the LHC are computed using  {\tt HIGLU
1.2}~\cite{higlu} for the $gg \to h/H/A$ process, including  the exact
contributions of the top and bottom quark loops at NLO--QCD  and the squark loops,
and the program {\tt bb@nnlo}  for $b\bar b \to h/H/A$ at NNLO-QCD. They are 
interfaced with  {\tt Suspect} for  the MSSM spectrum and  {\tt HDECAY} for  the
Higgs decay branching ratios.  The Higgs production cross sections and the
branching fractions for decays into $b \bar b$,  $\gamma \gamma, WW$ and $ZZ$ from
{\tt HIGLU} and {\tt HDECAY} are compared to those predicted by {\tt FeynHiggs}.
In the SM  both the $gg \to H_{SM}$ cross section and the branching fractions
agree within  $\sim 3\%$. Significant differences are  observed in the SUSY case,
with {\tt HDECAY} giving values of the branching fractions to $\gamma \gamma$ and
$WW$, $ZZ$ which  are on average 9\% lower and 19\% larger than those of {\tt
FeynHiggs} and have an r.m.s. spread of the distribution of the  relative
difference between the two programs of  18\% and 24\%, respectively~\cite{Arbey:2011aa}.

\subsection{Constraints}

We apply constraints from flavour physics, anomalous muon magnetic
moment, dark matter constraints and SUSY searches at LEP and the
Tevatron. These have been discussed in details in Ref.~\cite{Arbey:2011un}.
In particular, we consider the decay $B_s \to \mu^+ \mu^-$, which can
receive  extremely large SUSY contributions at large $\tan\beta$. An
excess of events in this channels has been reported  by the CDF-II
collaboration at the Tevatron~\cite{Aaltonen:2011fi} and upper limits by
the LHCb~\cite{Aaij:2012ac} and CMS~\cite{Chatrchyan:2012rg}
collaborations at LHC. Recently the LHCb collaboration has presented
their latest result for the search of this decay based on 1~fb$^{-1}$ 
of data. A 95\% C.L. upper limit on its branching fraction  is set at $4.5 \times
10^{-9}$~\cite{Aaij:2012ac}. After accounting for theoretical
uncertainties, estimated at the  11\% level \cite{Akeroyd:2011kd}
the constraint 
\beq
{\rm BR}(B_s \to \mu^+ \mu^-) < 5 \times 10^{-9}
\eeq
is used in this analysis. For large values of $\tan\beta$, this decay can be
enhanced by several orders of magnitude so that strong constraints on the
scalar contributions can be derived \cite{Hurth:2012jn}, and the small $M_A$ and
large $\tan\beta$ region can be severely constrained. As already remarked in
Ref.~\cite{Arbey:2011un}, the constraints obtained are similar and complementary
to those from the dark matter direct detection limits of XENON-100
\cite{Aprile:2011hi} and searches for the $A \to \tau^+ \tau^-$  decay.

Concerning the relic density constraint, we impose the upper limit derived
from the WMAP-7 result \cite{Komatsu:2010fb}
\begin{equation}
  10^{-4} < \Omega_{\chi} h^2 < 0.155\;,
\end{equation}
accounting for theoretical and cosmological uncertainties
\cite{Arbey:2008kv}.

The searches conducted by the ATLAS and CMS collaborations on the $\sqrt s=7$~TeV
data for channels with missing $E_T$~\cite{Aad:2011ib,Chatrchyan:2011zy} have
already provided a number  of constraints relevant to this study. These have
excluded a fraction of the  pMSSM phase space corresponding to gluinos below $\sim
600$~GeV and scalar quarks of the first  two generations below $\sim 400$~GeV.
These constraints are included using the same analysis discussed in
Ref.~\cite{Arbey:2011un},  extended to an integrated luminosity of 4.6~fb$^{-1}$.

Then, searches for the MSSM Higgs bosons in the channels $h/H/A  \to \tau^+ 
\tau^-$~\cite{Aad:2011rv,Chatrchyan:2012vp} have already excluded a significant
fraction of the $[M_A, \tan \beta]$ plane at low  $M_A$ values, $M_A \lsim 200$
GeV and $\tb \lsim 10$, and  larger values of $\tan \beta$ for $M_A \gsim 200$
GeV. These constraint on the pMSSM parameter space are already important. It is
supplemented by the search of light charged  Higgs bosons in top decays, $t \to
bH^+ \to b \tau \nu$, performed by the ATLAS collaboration  \cite{Aad:2011hh}
which is effective at low $M_A$ values, $M_A \lsim 140$~GeV, corresponding to
$M_{H^\pm} \lsim 160$~GeV.   

Following the Higgs discovery  at the LHC, the lightest Higgs boson in our
analysis  is restricted to have a mass in the range allowed by the results
reported by ATLAS and CMS:
\beq  
123~{\rm GeV} \leq M_h \leq 129~{\rm GeV} \label{Mh-range}  
\eeq
where the range is centred around the value corresponding to the average of the
Higgs mass values  reported by ATLAS and CMS, $M_h \simeq  126$ GeV,  with the
lower and upper limits accounting for the parametric  uncertainties from the SM
inputs given in eq.~(\ref{SM-inputs}), in particular the top quark mass,  and
the theoretical uncertainties in the determination of the $h$ boson mass. It is
also consistent  with the experimental exclusion bounds.

The impact of the Higgs mass value and its decay rates on the parameters of the pMSSM 
can be estimated by studying the compatibility of the pMSSM points with 
the first results reported by ATLAS \cite{ATLAS:2012zz} and CMS
\cite{CMS:2012zz} at the LHC and also by the Tevatron experiments
\cite{Tevatron:2012zz}. Starting from our  set of $6\times 10^7$ pMSSM points 
which are pre-selected for compatibility with the constraints discussed 
above, we consider the two decay channels giving the  Higgs boson evidence at
the LHC, $\gamma \gamma$ and $ZZ$ and include also the $b \bar b$ and $\tau \tau$ 
channels. In the following, we use the notation $R_{XX}$ to indicate the Higgs decay 
branching fraction to the final state $XX$, BR($h \rightarrow XX$), normalised to its 
SM value. We also compute the ratios of the product of production cross
sections  times branching ratios for the pMSSM points to the SM values, 
denoted by $\mu_{XX}$ for a given $h\to XX$ final state,
$\mu_{XX} = \frac{\sigma (h) \times {\rm BR}(h \to XX)}{\sigma (H_{\rm SM}) 
 \times {\rm BR} (H_{\rm SM}  \to XX)}$.
These are compared to the experimental values. For the $\gamma \gamma$, 
and $ZZ$ channels we take a weighted average of the results just reported by the experiments, 
as given in Table~\ref{tab:input} with their estimated uncertainties.

\begin{table}[!h]
\begin{center}
\begin{tabular}{|c|c|c|}
\hline
Parameter & Value & Experiment \\ \hline \hline
$M_H$     & 125.9$\pm$2.1 GeV & ATLAS \cite{ATLAS:2012zz} + CMS \cite{CMS:2012zz} \\ 
$\mu_{\gamma \gamma}$ & 1.71$\pm$0.33 & ATLAS \cite{ATLAS-2012-091} + CMS \cite{CMS-12-015} \\
$\mu_{Z Z}$ & 0.95$\pm$0.40 & ATLAS \cite{ATLAS-2012-092} + CMS \cite{CMS-12-016} \\
$\mu_{b \bar b}$ & $<$1.64 (95\% C.L.) & CMS \cite{CMS-12-019}\\ \hline
$\mu_{\tau \tau}$ & $<$1.06 (95\% C.L.) & CMS \cite{CMS-12-018}\\ \hline
\end{tabular}
\end{center}
\caption{Input parameters used for the pMSSM study.}
\label{tab:input} 
\end{table}
While the results are compatible with the SM expectations within the present accuracy, they highlight a possible enhancement in the 
observed rates for the $\gamma \gamma$ channel,  where ATLAS and CMS obtain $\mu_{\gamma \gamma}$ = 1.9$\pm$0.5 and 1.56$\pm$0.43, 
respectively. In the following, we do not take into account the theoretical uncertainties in the production cross section, which are 
estimated significant for the main production channel, $gg\to h$~\cite{THU,JB}. 

\subsection{The decoupling regime} 

\begin{figure}[h!] 
\centerline{\includegraphics[width=10cm]{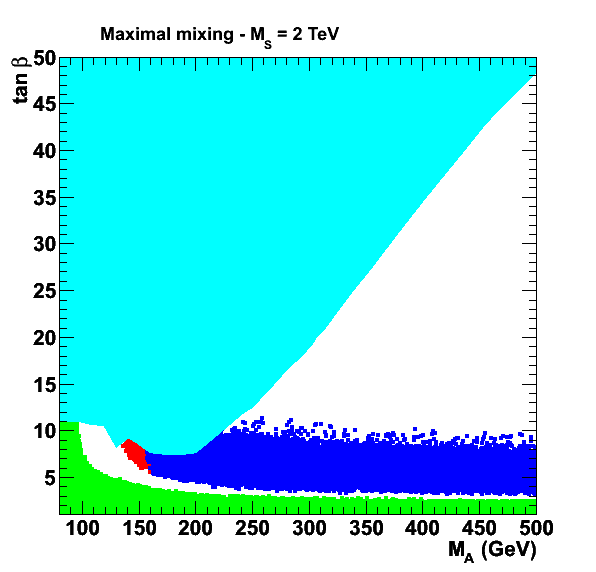}} 
\vspace*{-2mm}
\caption{\small The parameter space $[M_A, \tb]$ for $M_S=2$ TeV in the maximal
mixing scenario with the individual constraints from LEP2 (green), CMS $\tau^+ 
\tau^-$ searches (light blue) and flavor physics (red) displayed. The area in 
which  123 $\le M_H\le 129$~GeV is also shown (dark blue).} 
\label{fig:2} 
\end{figure}

Figure~\ref{fig:2} presents the parameter space $[M_A, \tb]$ in our benchmark
scenario with $M_S=2$~TeV in the maximal mixing scenario. The regions excluded  by
the various constraints  that we have imposed are indicated. The green area
corresponds to the   non-observation of Higgs bosons at LEP2 which excludes
$\tb\lsim 3$ at moderate to large $M_A$ values, $M_A \gsim 150$~GeV, but up to
$\tb \approx 5$--10 at low $M_A$ values.  The blue area is the  one  ruled out by
the latest published results of the CMS collaboration on the search of resonances
decaying into $\tau^+\tau^-$ final states; it touches the LEP2 band at small
$M_A$, but reduces in size when $M_A$ is increased. The small visible area in red
is the one excluded by the $B_s \to \mu^+ \mu^-$ constraint but, in fact, part of
the excluded  region is hidden  by the CMS blue area. 

To that, we superimpose the area in which we make the requirement 123 $\le
M_H\le 129$~GeV, that is indicated in dark blue. This band covers the entire range
of $M_A$ values and leaves only   the $\tan\beta$ values that are comprised
between  $\tb \approx 3$--5 and $\tb \approx 10$. Between the LEP2 and the
``$M_h$" blue band, one has $M_h <123$~GeV, while above the $M_h$ band, one has
$M_h > 129$~GeV and both areas are excluded. The requirement that the $h$ boson
mass should have the value measured at the LHC, even with the large uncertainty
that we assume,  provides thus a strong constraint on the $[M_A, \tb]$ parameter
space in the pMSSM.  

\begin{figure}[h!]
\begin{center}
\hspace*{-.5cm}\includegraphics[width=6.cm]{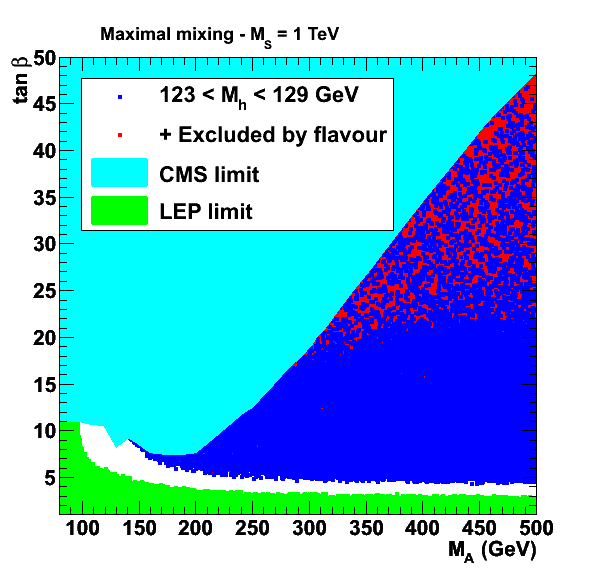}\includegraphics[width=6.cm]{maximalmix_MS2.png}\includegraphics[width=6.cm]{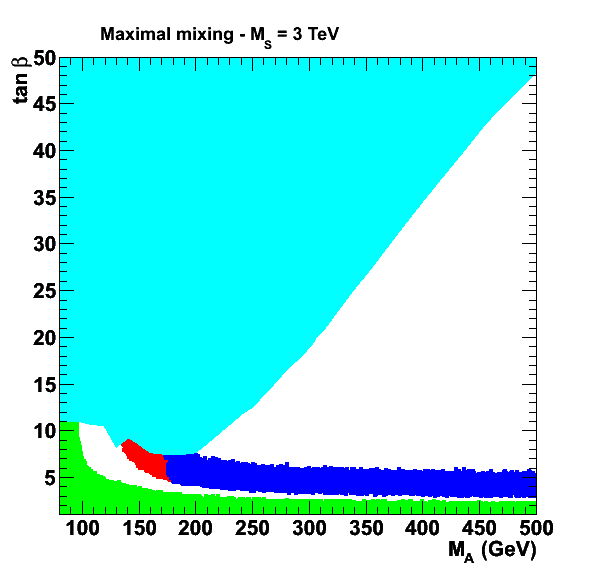}\\[.2mm]
\hspace*{-.5cm}\includegraphics[width=6.cm]{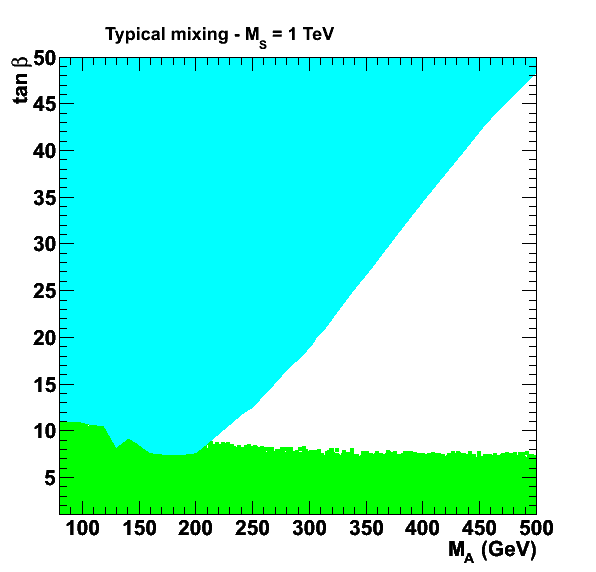}\includegraphics[width=6.cm]{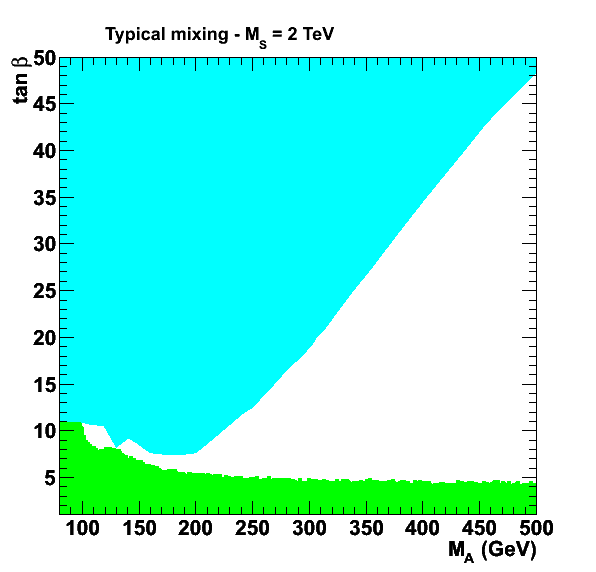}\includegraphics[width=6.cm]{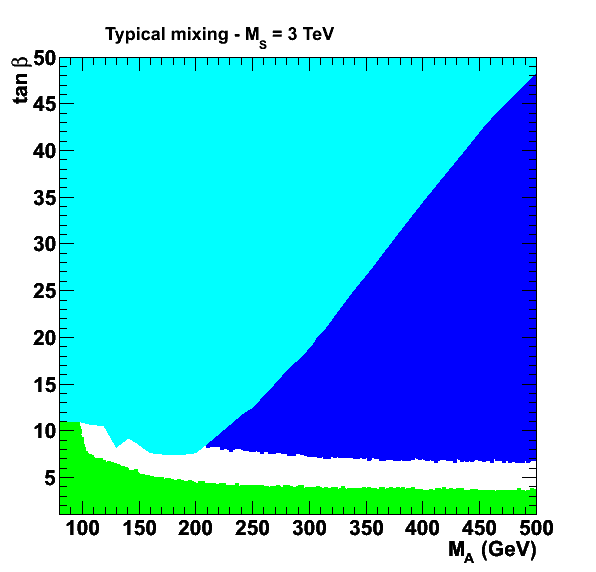}\\[.2mm]
\hspace*{-.5cm}\includegraphics[width=6.cm]{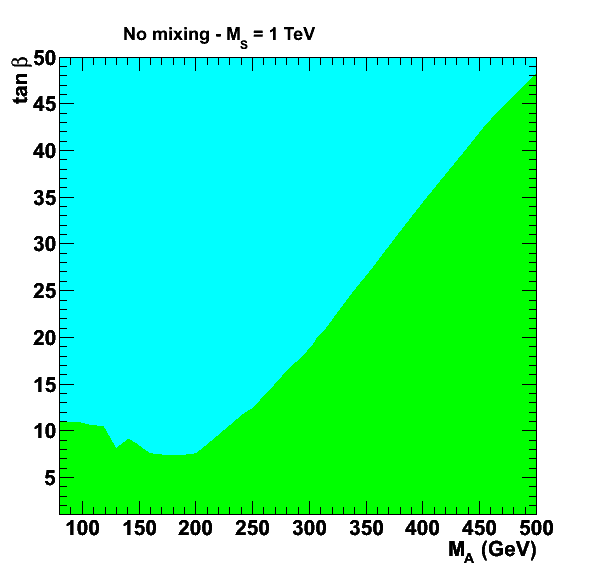}\includegraphics[width=6.cm]{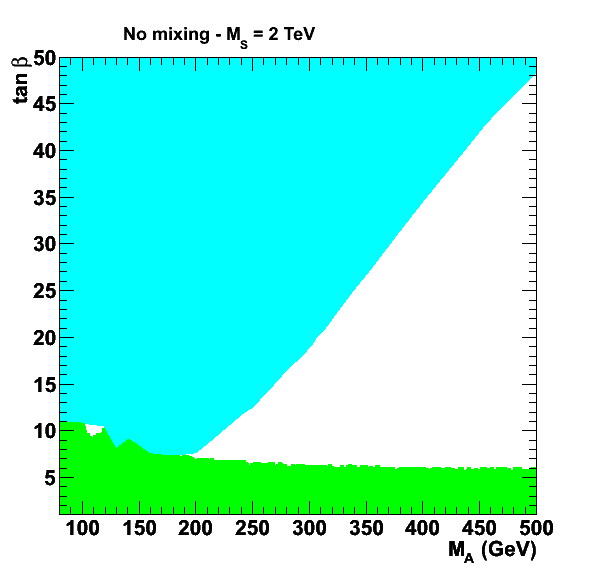}\includegraphics[width=6.cm]{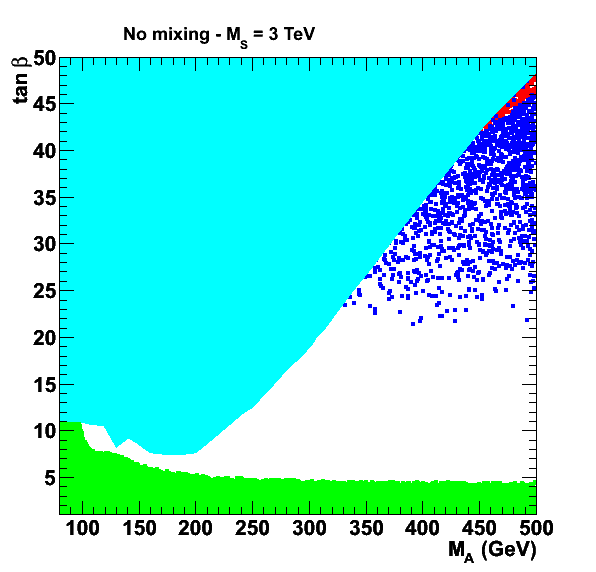}
\end{center}
\vspace*{-5mm}
\caption{\small The $[M_A,\tan \beta]$ plane for $M_S=1$,  2 and 3~TeV and for 
zero, typical and maximal mixing. The colour coding for the different regions is the same as in Fig.~\ref{fig:2}.}
\label{fig:3}
\end{figure}

In Figure~\ref{fig:3}, we show the same $[M_A, \tb]$  plane but for different
SUSY--breaking scales, $M_S=1,2$ and 3 TeV and for the zero, typical and maximal
mixing scenarios defined in eqs.~(4--6).  As can be seen, the situation changes
dramatically depending on the chosen scenario.   Still, in the maximal mixing
scenario with $M_S=3$~TeV the size of the $M_h$ band is reduced from above, as in
this case, already values $\tb \gsim 5$ leads to a too heavy $h$ boson, $M_h \gsim
129$ GeV. In turn, for $M_S= 1$~TeV, the entire space left by the LEP2  and CMS
Higgs constraints is covered with many points at $\tb \gsim 20$ excluded by the
flavor constraint.   Nevertheless, the possibility with $M_S\approx 1$ TeV will
start to be challenged by the search for squarks at the LHC when 30~fb$^{-1}$ of
data will be collected by the experiments. In the no--mixing scenario, it is
extremely hard to obtain a Higgs mass of $M_h \geq 123$~GeV and all parameters
need to be maximised: $M_S=3$~TeV and $\tb \gsim 20$; a small triangle is thus
left over, the top of which is challenged  by the flavor constraints. The typical
mixing scenario resembles to the no--mixing scenario, with the notable difference
that for $M_S=3$~TeV, the entire space not excluded by the  LEP2 and CMS
constraints allow for an acceptable value of $M_h$. 

In the discussion so far, we have adopted the value $m_t$ = (173$\pm$1)~GeV for 
the top quark mass as  measured by the CDF and D0 experiments at the
Tevatron~\cite{Lancaster:2011wr}.  This implicitly assumes that this mass corresponds 
to the top quark pole mass, i.e. the mass in the  on--shell scheme, which serves as
input in the calculation of the radiative corrections in the pMSSM Higgs sector 
and, in particular, to the  mass $M_h$. However, the mass measured at the Tevatron is not 
theoretically well defined and it is not proved that it corresponds indeed to the pole mass 
as discussed in~\cite{pole-mt}.  For an unambiguous and well-defined determination of the
top quark mass, it is appears to be safer to use the value obtained from the determination of the 
top quark pair production cross section measured at the Tevatron, by comparing the
measured value with the theoretical prediction at higher orders. This determination has been
recently performed yielding the value of (173.3$\pm$2.8)~GeV~\cite{pole-mt}
for $m_t^{\rm{pole}}$. The central value is very close to that measured 
from the event kinematics but its uncertainty is larger as a result of the experimental 
and theoretical uncertainties that affect the measurement.

\begin{figure}[h!]
\vspace*{-4mm}
\hspace*{-.5cm}\includegraphics[width=6.2cm]{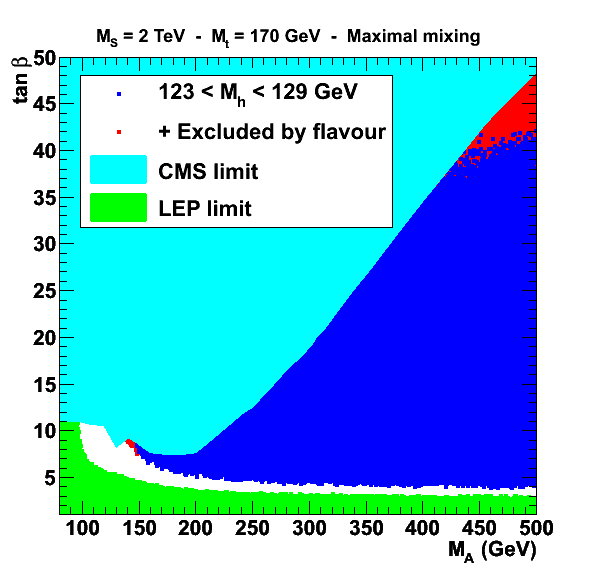}\includegraphics[width=6.cm]{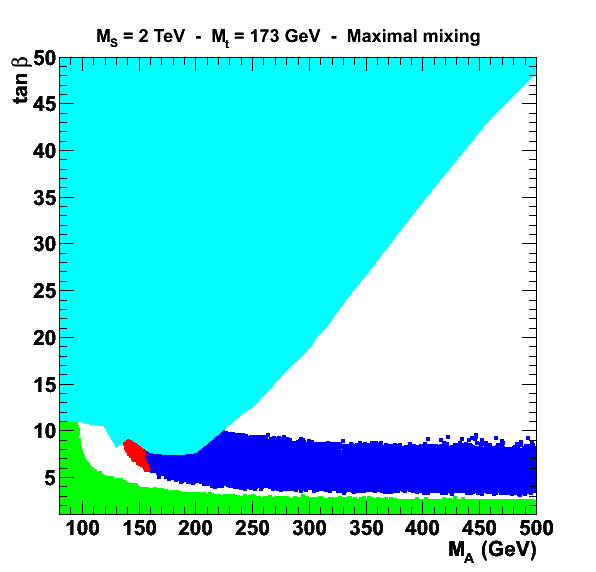}\includegraphics[width=6.2cm]{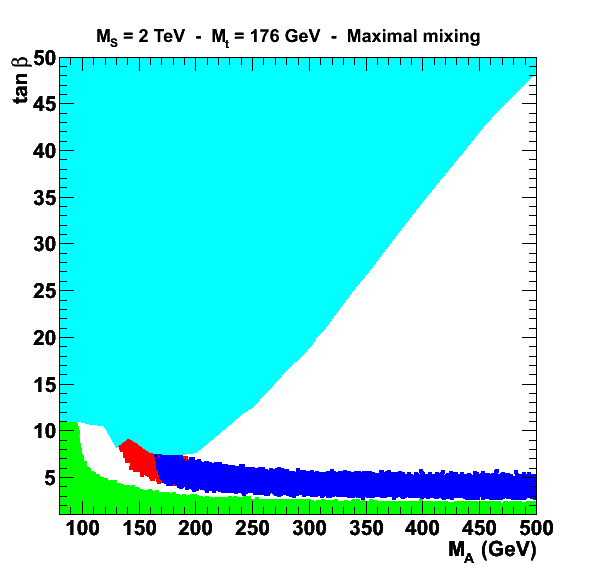}
\vspace*{-7mm}
\caption{\small The parameter space $[M_A,\tan \beta]$  for $M_S=2$ TeV,
maximal mixing  and 
three values of the top quark mass $m_t= 170$~GeV (left), 173 GeV (centre) and
176~GeV (right).}
\label{fig:4}
\end{figure}

\begin{figure}[h!]
\centerline{
\includegraphics[width=10.cm]{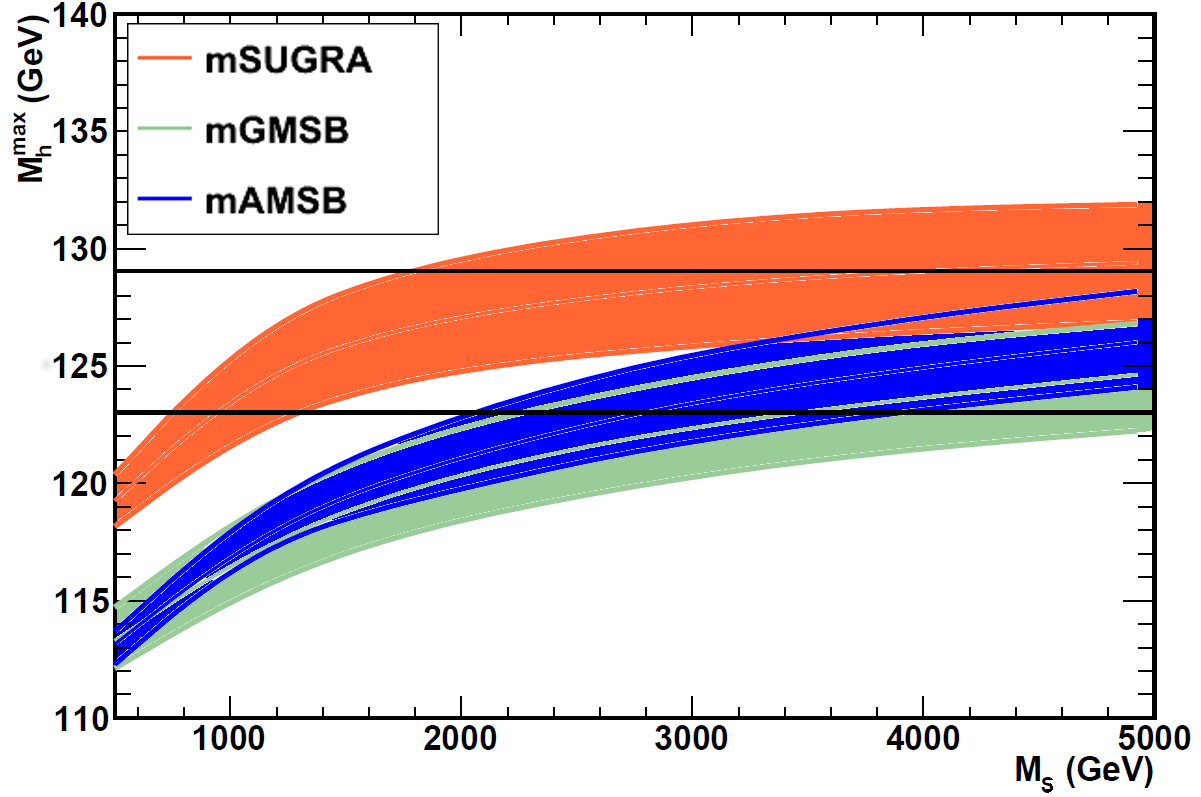}
}
\vspace*{-2mm}
\caption{\small Maximal Higgs mass in the constrained MSSM scenarios
mSUGRA, mAMSB and mGMSB, an a function of the scale $M_S$ when the top 
quark mass is varied in the range $m_t=\;$170--176 GeV.}
\label{fig:5}
\end{figure}

It is interesting to assess the impact of a broader mass range for the top quark. We return to our benchmark 
scenario with $M_s=2$~TeV and  maximal stop mixing and draw the ``$M_h$" bands using the top quark mass 
values of 170~GeV and 176~GeV corresponding to the wider uncertainty interval quoted above. The result 
is shown in Figure~\ref{fig:4}. A 1~GeV change in $m_t$ input leads to a $\sim$1~GeV change in the corresponding 
$M_h$ value. The smaller value of $m_t$  would open up more parameter space as the region in which $M_h \gsim 129$~GeV  
will be significantly reduced. In turn, for $m_t=176$ GeV, the corresponding $h$ boson mass increases 
and the dark--blue area quite significantly shrinks, as a result.  It must be noted that for $m_t=170$
GeV,  the no--mixing scenario  would be totally excluded for $M_S \lsim 3$~TeV, while in the typical 
mixing scenario  only a small area at high $ \tb$ will remain viable. For $m_t=176$~GeV significant 
$[M_A, \tb]$ regions that was excluded when taking the $\pm$~1~GeV uncertainty for top mass value becomes 
allowed. 
The impact of the value of $m_t$ is thus extremely significant. This is even more
true in constrained scenarios, where the top mass also enters in the
evaluation of the soft SUSY--breaking parameters and the minimisation of the
scalar potential. To  visualise the impact of $m_t$, we have repeated the study 
presented in Ref.~\cite{ABDMQ}, presenting the maximal $M_h$ value reached  when 
scanning over all the parameters of  the  minimal SUGRA, AMSB and GMSB models. 
Figure~\ref{fig:5} shows the result with the $M_h^{\rm max}$ value as a function of 
$M_S$  taking $m_t$=173$\pm$3~GeV. While for $m_t$ = 173~GeV, there is no region of the 
parameter space of the mAMSB and mGMSB models which satisfies $M_h \gsim$ 123~GeV, 
for $M_S\lsim 3$ TeV assumed in~\cite{ABDMQ}, and the models are disfavoured,  using 
$m_t$=176~GeV, the regions of these mAMSB and mGMSB models beyond $M_S=2$ TeV become 
again viable. This will be also the case of some of the variants and even more constrained 
mSUGRA scenarios.  Further, even for $m_t$=173~GeV,  if we move the $M_S$ upper limit from the 3~TeV boundary 
adopted in Ref.~\cite{ABDMQ} to $M_S=5$ TeV, these models have region of their parameters compatible with 
the LHC Higgs mass. 

\begin{figure}[h!]
\vspace*{-3mm}
\hspace*{-.2mm}\includegraphics[width=7.cm]{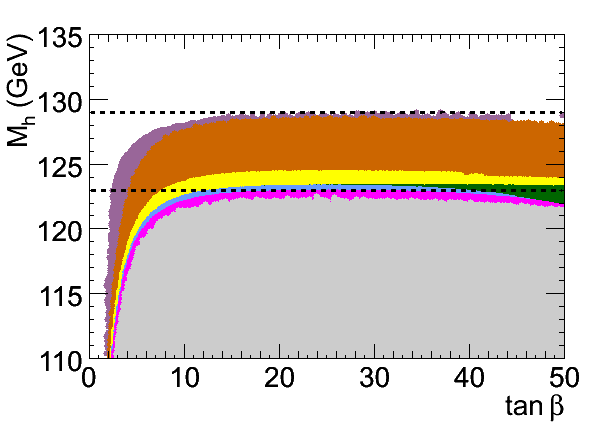}~~~\includegraphics[width=2.cm]{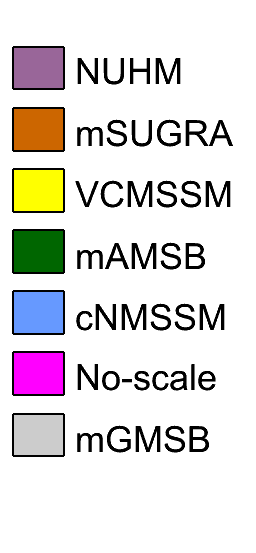}~~~\includegraphics[width=7.cm]{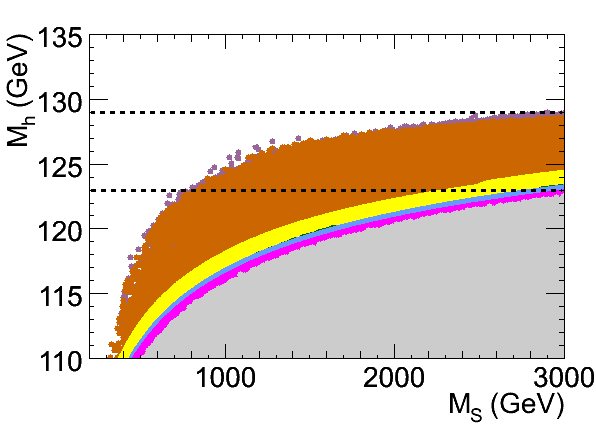}
\vspace*{.5mm}
\caption{\small The maximal $h$ mass value $M_h^{\rm max}$ as functions of 
$\tb$ (left) and $M_S$ (right) in the mASMB, mGMSB as well as in mSUGRA and 
some of its variants.  The basic parameters of the models are varied within the
ranges given in Ref.~\cite{ABDMQ}; the top quark mass is fixed to
$m_t=173$ GeV.}
\label{fig:5b}
\end{figure}

Finally, we comment on the impact of increasing
the $M_h$ allowed range from 123 GeV $\leq M_h \leq 127$ GeV as was done in
Ref.~\cite{ABDMQ} relying on the 2011 LHC data, to the one adopted here, 123 GeV
$\leq M_h \leq 129$ GeV, in the various constrained models discussed in that
reference (and to which we refer for the definition of the models and for the 
ranges of input parameters that have been adopted). The outcome is shown in
Fig.~\ref{fig:5b} where the maximal $h$ mass  value obtained by scanning the
basic input parameters of the model over the appropriate ranges. In the
left--hand side, $M_h^{\rm max}$ is displayed as a function of $\tb$ and in the
right--hand side as a function of $M_S$. As the lower bound $M_h^{\rm max} \geq
123$ GeV  is the same as in our previous analysis, the mASMB, mGMSB and some
variants of the mSUGRA model such as the constrained NMSSM (cNMSSM), the
no-scale model and the very constrained MSSM (VCMSSM) scenarios are still
disfavoured. However, for mSUGRA and the non--universal Higgs mass model (NUHM),
all values of $\tb\gsim 3$ and 1 TeV $\lsim M_S \lsim 3$ TeV lead to an
appropriate value of $M_h$ when including  the uncertainty band.

\subsection{The other (non--SUSY) regimes}

\begin{figure}[!h]
\begin{center}
\vspace*{-2mm}
\includegraphics[width=10.cm]{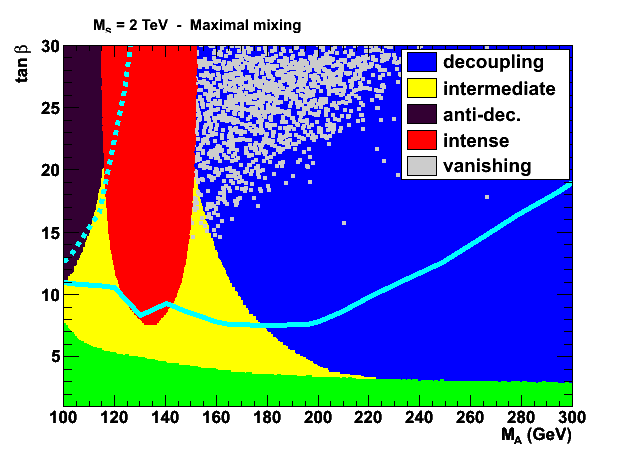}
\end{center}
\vspace*{-5mm}
\caption[]{\small Parameter space for the various regimes of the MSSM Higgs
sector as defined in the text and in eq.~(8) in the $[M_A,\tb]$ plane,  in the
maximal  mixing scenario  with $M_S=2$ TeV. The upper limit constraints from $\Phi \to \tau
\tau$ (continuous light  blue line)  and $t \to H^+ b$ (dashed blue line)
searches  at the LHC are shown together with the LEP2 excluded region (green area).}
\label{fig:regimesexp}
\end{figure}

The other regimes of the pMSSM Higgs sector, apart from the
decoupling and the SUSY regimes, occur for low to intermediate values of the pseudoscalar 
Higgs mass,  $M_A \lsim 200$ GeV, and relatively large $\tb$ values.  These are the anti--decoupling,  
the intense, the intermediate and the vanishing coupling regimes.
These are constrained by the results of the LEP2 and LHC searches. The LEP2 results for
$M_A \lsim 200$ GeV and not too large $M_S$ values, lead to  $\tb \gsim 3, 8$
and  10 for, respectively the maximal, typical and no--mixing scenarios; see
Fig.~\ref{fig:3}. The negative search for Higgs particles
in $\tau$-lepton final states, $pp \to \Phi \to \tau^+\tau^-$, by the ATLAS and
CMS collaborations places further constraints.  While in the decoupling regime, the relevant Higgs states
would be $\Phi=A+H$, these are $\Phi=A+h$ and $\Phi=A+H+h$ in the
anti-decoupling and intense coupling regimes, respectively. As already
mentioned,  one would have in the three regimes the same signal cross section
times branching ratios $\sigma(pp \to \Phi \to \tau^+\tau^-) \approx 
\sigma(b\bar b+gg \to A)\times {\rm BR}(A \to \tau^+ \tau^-)$ almost
independently of the mixing scenario and the other pMSSM parameters
\cite{JB}. The constraint from the CMS published results alone with the
$\approx 5$ fb$^{-1}$ of data collected in 2011 \cite{Chatrchyan:2012vp} imposes
$\tb \gsim 10$, as shown in  Fig.~\ref{fig:regimesexp} where we zoom on
the plane $[M_A, \tb$] at low to intermediate $M_A$ values, for the
maximal mixing scenario and $M_S=2$ TeV.  

This limit can be strengthened by the same $\tau^+\tau^-$ search performed by
the ATLAS collaboration \cite{Aad:2011rv} and also by the $t \to H^+ b$ search
in top decays\cite{Aad:2011hh} which is effective for
$M_A=\sqrt{M_{H^\pm}^2+M_W^2} \lsim 130$ GeV and which, as can be seen in
Fig.~\ref{fig:regimesexp},  excludes large $\tb$ values for which BR$(t \to b
H^+)$ is significant. Put together, these constraints  exclude entirely
both the anti-decoupling and intense coupling  regimes. Would remain then, the
intermediate coupling regime with $\tb \approx 5$--8 when the LEP2 constraint is
also imposed. Depending on the mixing  scenario, most of it will be excluded 
by the $M_h\approx 126$ GeV constraint (see Fig.~\ref{fig:3}).

A very interesting possibility would be that the observed Higgs particle at the
LHC is actually the $H$ state, while the lighter $h$ boson has suppressed
couplings to $W/Z$ bosons and top quarks, allowing it to escape  detection.  In
this case,  the $H$ couplings to bottom quarks should not be too enhanced, $\tb
\lsim 8$, not to be in conflict with the $\tau^+\tau^-$ and $t \to bH^+$  searches
above. For the $H$ boson to be SM--like, one should have $M_A \approx M_H \approx 
126$ GeV and not too low $\tb$ values, $\tb \gsim 7$--10. One is then in the
borderline between  the  anti-decoupling and the intermediate coupling regimes. 
We have searched for points in which indeed $M_H \approx 126$ GeV with couplings
to $VV$ states, $g_{HVV} \gsim 0.9$, such that the $H\to ZZ$ and  $H\to \gamma
\gamma$  (which mainly occurs through a $W$-boson loop) decays are not suppressed
compared to the measured values by ATLAS and CMS given in Table~\ref{tab:input}. In our scan,
out of the $10^6$ points, before imposing any LHC--Higgs constraint, only $\approx 20$ 
points fulfilled the above requirements. These points are then completely excluded 
once the flavor constraints, in particular those from the $b\to s\gamma$ radiative decay, 
are imposed. Hence, the possibility that the observed Higgs particle at the  LHC is 
not the lightest $h$ particle appears highly unlikely according to the result of our 
scan of the parameter space. Combining the $h/A \to \tau^+\tau^-$ and the $t \to bH^+$ 
constraints and including the results on the new 8~TeV data should further constrain 
the parameter space and completely exclude this scenario. 

Finally, the vanishing coupling regime is strongly disfavoured by the LHC and
Tevatron data that are summarised in Table~\ref{tab:input}.  The observation  of $H\to ZZ$ final
states by the ATLAS and CMS  collaborations rules out the possibility  of
vanishing $hVV$ couplings. The reported excess of events in the $q\bar q \to VH
\to Vb\bar b$ process by the CDF and D0 collaborations  seem also to rule out both
the  vanishing $hbb$ and $hVV$ coupling scenarios. However, there is still the
possibility that these couplings are smaller than  those predicted in the SM case,
in particular because of the effects of SUSY particles at high $\tb$
\cite{vanishing}. We are then in the SUSY--regime to which we turn now. 

\subsection{The SUSY regime}

In the SUSY regime, both the Higgs production cross section in gluon--gluon
fusion  and  the Higgs decay rates can be affected by the contributions of SUSY
particles. This makes a detailed study of  the pMSSM parameter space in relation
to the first results reported by the ATLAS and CMS collaborations especially
interesting for its sensitivity  to specific regions of the pMSSM
parameter space. In particular, the branching fraction for the  $\gamma \gamma$
decay of the $h$ state  is modified by Higgs mixing effects outside the decoupling
regime as was discussed above, by a change of the $hbb$ coupling  due to SUSY
loops \cite{vanishing}, by light superparticle contributions to the
$h\gamma\gamma$ vertex~\cite{ST-gg,ST-pp,Carena:2011aa} and by invisible $h$
decays into light neutralinos \cite{H-LSP}. 

We study these effects on the points of our pMSSM scan imposing  the LHC 
results as constraints. The numerical values adopted in the analysis are 
given in Table~\ref{tab:input}, assuming in the following on that the observed 
particle is the $h$ state. 
First, we  briefly summarise the impact of the SUSY particles on the Higgs decay  
branching fractions, staring from invisible decays, and production cross sections. 
Then we discuss our finding on the impact of the LHC and Tevatron data on the pMSSM 
parameters.

\subsubsection{Invisible Higgs decays}

Despite the fact that the discovered particle has a sufficient event rate in visible 
channels to achieve its observation, it is interesting to consider the regions of 
parameter space in which invisible Higgs decays occur. This scenario has recently been 
re-considered in~\cite{AlbornozVasquez:2011aa,Arbey:2011aa}. 
Besides the value of $M_h$, the invisible branching ratio 
BR($h \to \chi_1^0 \chi_1^0)$ is controlled by four parameters: the gaugino 
masses $M_1$ and $M_2$, the higgsino parameter $\mu$ and $\tb$. They enter the $4
\times 4$ matrix $Z$ which diagonalises the neutralino mass matrix. They also
enter the Higgs coupling to neutralinos which, in the case of the LSP, is  
\beq
g_{h \chi_1^0 \chi_1 0} \propto  \left( Z_{12}- \tan\theta_W Z_{11} \right)  
\left(\sin\beta Z_{13} + \cos\beta Z_{14} \right)
\label{ghchi}
\eeq
if we assume the decoupling limit not to enhance the $h\to  b\bar b$ channel
which would significantly reduce the invisible decay. In this coupling, $Z_{11},
Z_{12}$ are the gaugino components and $Z_{13},Z_{14}$ the higgsino components.
Thus, the coupling vanishes if the LSP is a pure  gaugino, $|\mu| \gg M_1$
leading to $m_{\chi_1^0} \approx M_1$, or  a pure higgsino, $M_1 \gg |\mu|$ with
$m_{\chi_1^0}\approx |\mu|$. 

\begin{figure}[h!]
\vspace*{-5mm}
\begin{center}
\begin{tabular}{cc}
\includegraphics[width=8.2cm]{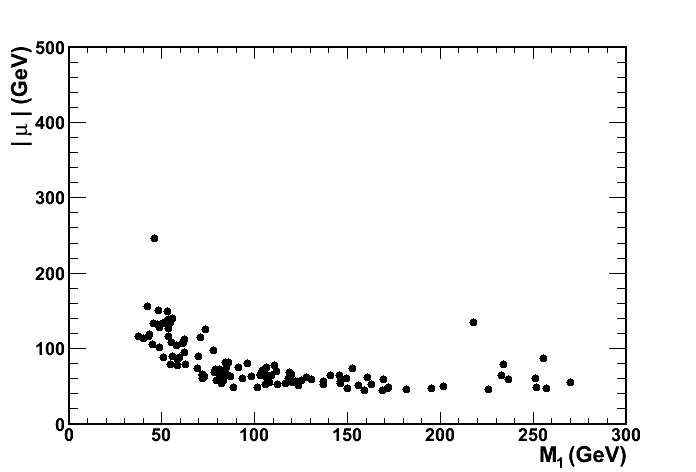} &
\hspace*{-0.75cm}\includegraphics[width=8.2cm]{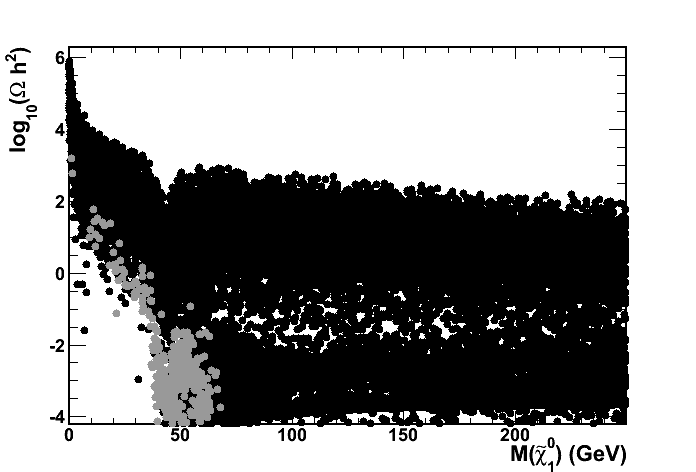} \\
\end{tabular}
\end{center}
\vspace*{-5mm}
\caption{\small Left: Points in the $[M_1, \mu$] parameter space where the invisible branching
fraction BR$(h\to \chi_1^0 \chi_1^0) \geq$ 0.15 from a pMSSM scan where we impose the LEP constraint on the 
$Z$ invisible width and neutralino relic density $\Omega_{chi} h^2$. Right: $\Omega_{\chi} h^2$ as a function of $m_{\chi^0_1}$ 
with all the selected pMSSM points in black and those giving a  BR$(h\to \chi_1^0 \chi_1^0) \geq$ 0.15 in grey.}
\label{fig:inv}
\end{figure}

For the invisible decay to occur, a light LSP, $m_{\chi_1^0} \le \frac12 M_h$ is required.
Since in the pMSSM, the gaugino mass universality $M_2 \approx 2M_1$ is  relaxed, one can
thus have a light neutralino without being in conflict  with data. 
The constraint from the $Z$ invisible decay width measured at LEP restricts the parameter 
space to points where the $\tilde\chi^0_1$ is bino-like, if its mass is below 45~GeV, and thus 
to relatively large values of the higgsino mass parameter $|\mu|$. Since a large decay width into 
$\tilde \chi^0_1 \tilde \chi^0_1$ corresponds to small values of $|\mu|$, this remove a large 
part of the parameter space where the invisible Higgs decay width is sizable. 
Still, we observe invisible decays for 45~GeV$< M_{\tilde\chi^0} < M_{h^0}/2$ and $| \mu | <$ 150, 
corresponding to a combination of parameters where the $\tilde\chi^0_1$ is a mixed higgsino-gaugino 
state~\cite{Arbey:2011aa}. These pMSSM points are shown in the $[M_1,\mu$] plane in the left panel of 
Fig.~\ref{fig:inv}.

If the LSP at such a low mass were to be the dark-matter particle, with the relic 
density given in eq.~(12),  it should have an efficient annihilation rate into SM 
particles.  The only possible way for that to occur would be $\chi^0_1 \chi^0_1$ 
annihilation through the $s$--channel light $h$   pole\footnote{The other possible channels
are strongly suppressed or ruled out. The co--annihilation with charginos,
heavier neutralinos and staus is not effective as these particles need to be
heavier than $\approx 100$ GeV and thus the mass difference with the LSP is too
large. The annihilation through the $A$--pole needs $M_A \approx 2m_{\chi_1^0}
\lsim M_h$ and sizable $\tan\beta$ values, which is the anti-decoupling regime
that is excluded as discussed above. Remains then the bulk region with staus
exchanged in the $t$--channel in $\chi^0_1 \chi^0_1 \to \tau^+  \tau^-$
(sbottoms are too heavy) which is difficult to enhance as the LSP is 
bino--like.} \cite{Hpole} which  implies that $m_{\chi_1^0}  \lsim \frac12 M_h$
to still have a non--zero invisible branching ratio, as shown in the right panel of Fig.~\ref{fig:inv}, 
where the pMSSM points satisfying BR($h \to \chi_1^0\chi_1^0) \geq 5\%$ are shown in the
plane $[m_{\chi_1^0},\log_{10} (\Omega h^2)]$.
However, because the partial decay  width $\Gamma(h \to \chi_1^0 \chi_1^0)$ is suppressed
by a factor $\beta^3$ near the $M_h \approx 2 m_{\chi_1^0}$ threshold, with the
velocity $\beta= (1- 4m_{\chi_1^0}^2/M_h^2 )^{1/2}$, the invisible branching
fraction is rather small if the WMAP dark matter constraint is to hold. MSSM light neutralinos 
compatible with claims of direct detection dark matter signals are also consistent with collider 
bounds~\cite{Arbey:2012na}.


\subsubsection{Sparticle effects on the $\mathbf{hb\bar b}$ coupling}

SUSY particles will contribute to the $hb\bar b$ coupling as there are additional
one--loop  vertex corrections that modify the tree--level Lagrangian that
incorporates  them \cite{vanishing}.  These corrections involve bottom
squarks and gluinos in the loops, but there  are also possibly large corrections
from stop and chargino loops. Both can be large since they grow as  $\mu
\tb$ or $A_t \mu \tb$ \cite{vanishing}  
\begin{eqnarray}
\Delta_b \approx \frac{2\alpha_s}{3\pi} \frac{ m_{\tilde{g}} \mu \tb}
{ {\rm max}(m_{\tilde{g}}^2, m_{\tilde{b}_1}^2, m_{\tilde{b}_2}^2)} +
\frac{m_t^2}{8\pi^2 v^2 \sin^2\beta} \frac{ A_t \mu \tb}{ {\rm max}
(\mu^2,m_{\tilde{t}_1}^2, m_{\tilde{t}_2}^2)} .
\label{Deltab} 
\eeq 
Outside the decoupling limit, the reduced $b\bar b$  couplings of the  $h$ state
are given in this case by 
\beq
 g_{hbb} \approx g_{Abb} \approx   \tb (1-\Delta_b)
\eeq
and can be thus
significantly reduced or enhanced\footnote{These corrections also affect the Higgs
production cross sections in the channels $gg+b\bar b \to \Phi$. However, in the
cross sections times branching ratios for the $\tau^+ \tau^-$  final states, they
almost entirely cancel as they appear in both the  production rate and the total
Higgs decay width~\cite{JB}.} depending on the sign of $\mu$ and, possibly, also
$A_t$. 
This is exemplified in the left panel of Fig.~\ref{fig:bbmutb}, where the ratio 
$R_{bb}\equiv {\rm BR}(h\to b\bar b)/{\rm BR}(H_{\rm SM} \to b\bar b)$ is shown 
as a function of the parameter $\mu\tan\beta$ before the constraints of 
Table~\ref{tab:input}. The two branches in the histogram are due to the 
sbottom and stop contributions in which $R_{bb}$ is increased or
decreased depending on the sign of $\mu$.  

\begin{figure}[h!]
\vspace*{-5mm}
\begin{center}
\begin{tabular}{cc}
\includegraphics[width=8.75cm]{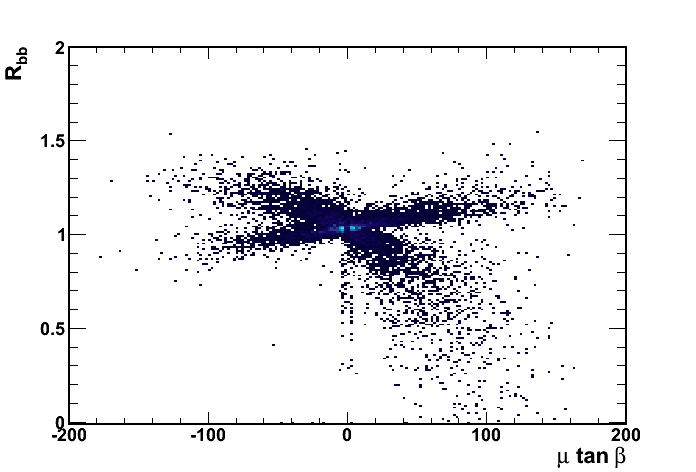} &
\hspace*{-1.10cm} \includegraphics[width=8.75cm]{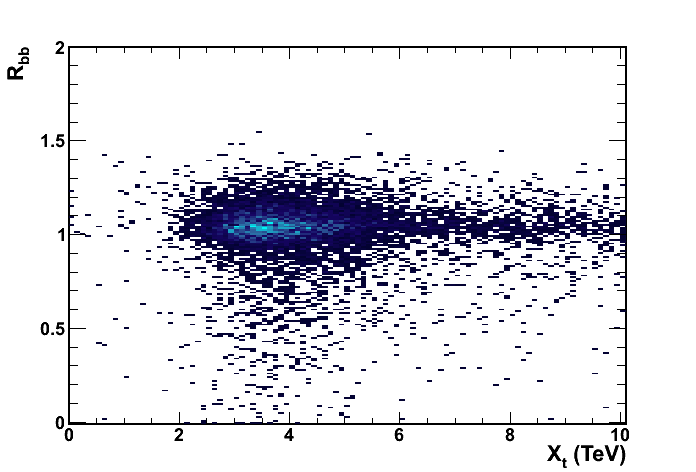} \\
\end{tabular}
\end{center}
\vspace*{-5mm}
\caption{\small  
(Left) $R_{bb}$ values for a sample of pMSSM points as a function of the  product of the $\mu \tan \beta$ showing the reduction at 
large values of $\mu \tan \beta$. The reduction in a narrow strip at small values of $\mu$ is due to decays into $\chi \chi$.  
(Right) the same as a function of $A_t$.}
\label{fig:bbmutb}
\end{figure}
\begin{figure}[h!]
\vspace*{-5mm}
\begin{center}
\includegraphics[width=8.75cm]{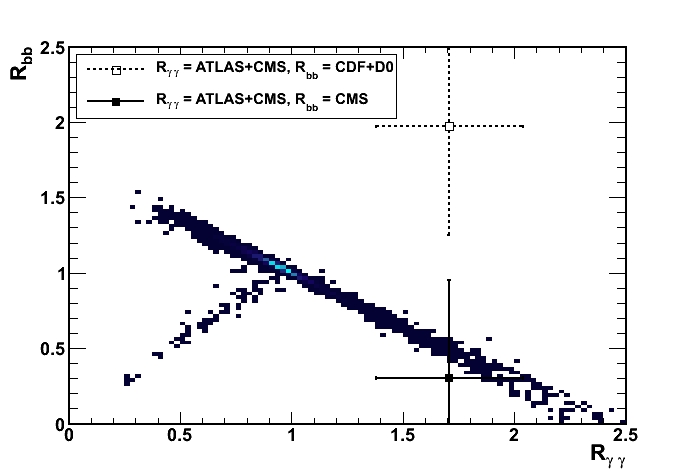} 
\end{center}
\vspace*{-5mm}
\caption{\small  
$R_{bb}$ as a function of $R_{\gamma \gamma}$, showing their 
anti-correlated  variation; the points corresponding to a decrease of both 
ratios are due to an enhancement of invisible decays to light neutralinos. The values
of $R_{\gamma \gamma}$ obtained by ATLAS+CMS and $R_{bb}$ corresponding to the CMS and CDF+D0 
searches are overlayed as comparison.}
\label{fig:bbgg}
\end{figure}
A  deviation of the partial $h\to b\bar b$ width will enter the total Higgs 
width, which is dominated by the $b \bar b$ channel, and change the $R$ values for
the different Higgs decay channels. A reduction of $R_{bb}$ would thus lead  to
an  enhancement of the $\gamma \gamma$  and the $WW$/$ZZ$ branching fractions.
Figure~\ref{fig:bbgg} shows the values of $R_{bb}$ and $R_{\gamma \gamma}$ in
which we  observe a highly anti-correlated  variation of the two ratios, with the
exception of the cases where the opening of the decay $h \to \chi \chi$ suppresses
both branching fractions. The preliminary results from LHC and the Tevatron are overlayed.

\subsubsection{Sparticle contributions to the hgg and h$\gamma\gamma$ vertices}

Scalar top quarks can alter significantly the $gg\to h$ cross section as well as the 
$h\to \gamma \gamma$ decay width~\cite{ST-gg}. The current eigenstates $\tilde
t_L , \tilde t_R$ mix strongly, with a mixing angle $ \propto m_t X_t$, so 
that for large $X_t=A_t-\mu/\tb$ values\footnote{On should assume $X_t$ values
such that $A_t \lsim 3M_S$ to avoid dangerous  charge and colour breaking minima.
In addition, if $X_t \gsim \sqrt 6 M_S$, the radiative corrections to the $h$
boson mass become small again and it would be difficult to attain the value $M_h
\approx 126$ GeV.}  there is a lighter mass eigenstate $\tilde t_1$ which can be 
much lighter than all other scalar quarks, $ m_{\tilde t_1} \ll M_S$. The
coupling  of the $h$ boson to the $\tilde t_1$ states in the decoupling regime
reads  
\beq 
g_{h \tilde{t}_1 \tilde{t}_1 } = \cos 2\beta M_Z^2 \left[ \frac{1}{2} \cos^2 
\theta_t - \frac{2}{3} s^2_W \cos 2 \theta_t \right] + m_t^2 +  \frac{1}{2} \sin
2\theta_t  m_t X_t \ , \  \sin {2\theta_t} = \frac{2 m_t X_t} {
m_{\tilde{t}_1}^2 -m_{\tilde{t}_2}^2 }  
\eeq 
In the no-mixing scenario $X_t\approx 0$, the coupling above is $\propto m_t^2$
and the scalar top contribution to the $hgg$ amplitude is small, being damped by a
factor $1/m_{\tilde t_1}^2$ and  interferes constructively with the top quark
contribution to increase the $gg\to h$ rate.  However, since in the no--mixing
scenario $M_S=\sqrt{m_{\tilde t_1} m_{\tilde t_2}}$ has to be very large for the
$h$ boson mass to reach a value $M_h \approx 126$ GeV, the stop contribution
to the $hgg$ vertex, $\propto m_t^2/M_S^2$, is very small. In the maximal mixing
scenario, $X_t \approx \sqrt 6 M_S$, it is the last component  of
$g_{h \tilde{t}_1 \tilde{t}_1 }$  which dominates and becomes very
large, $\propto - (m_tX_t/m_{\tilde t_2})^2$. However, in this case, the  large
contribution of a light stop to the $hgg$ amplitude interferes destructively
with the top quark contribution  and the $gg \to h$ cross section is suppressed.
For $m_{\tilde t_1}\approx 200$ GeV and $X_t \approx 1$ TeV, we obtain a
factor of two smaller $gg\to h$ rate.  In the case of sbottom squarks, the same
situation may occur for large sbottom mixing $X_b=A_b-\mu
\tb$. However, for large value of $M_S$, it is more difficult to obtain  a small
enough $m_{\tilde b_1}$ state to significantly affect the $gg\to h$ cross
section. 

In the case of the $h \gamma \gamma$ decay amplitude, there is the additional SM
contribution  of the $W$ boson, which is in fact the dominant. Also, it has
the opposite sign to that from the top quark and, hence, when stops are light and
have a strong mixing, they will tend to increase the $h\gamma\gamma$ amplitude.
However,  because the $W$ contribution is by far the largest, the stop impact
will be much more limited compared to the $ggh$ case and we can expect to
have only a $\approx 10\%$ increase of the $h\to \gamma \gamma$ decay rate for
$m_{\tilde t_1}\approx 200$ GeV and $X_t \approx 1$ TeV
\cite{ST-gg,ST-pp,villadoro}. Therefore, for light and strongly  mixed stops,
the cross section times branching ratio $\mu_{\gamma \gamma}$ is always smaller 
unity  and relatively light stops do not entail an enhancement of the $\gamma\gamma$ 
yield.  
\begin{figure}[h!]
\begin{center}
\vspace*{-4mm}
\begin{tabular}{cc}
\includegraphics[width=8.0cm]{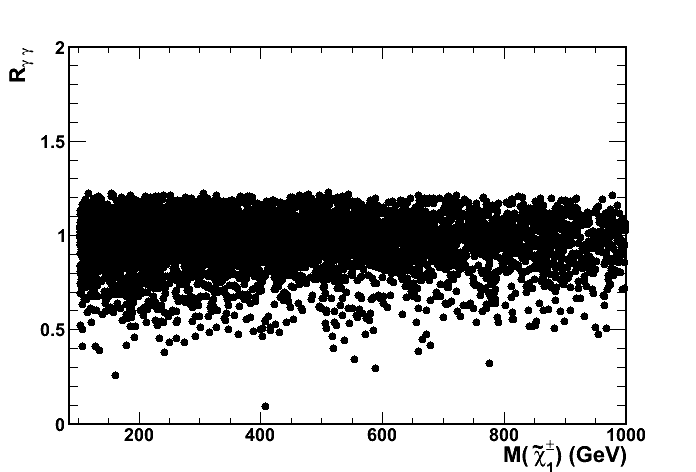} &
\hspace*{-0.75cm} \includegraphics[width=8.0cm]{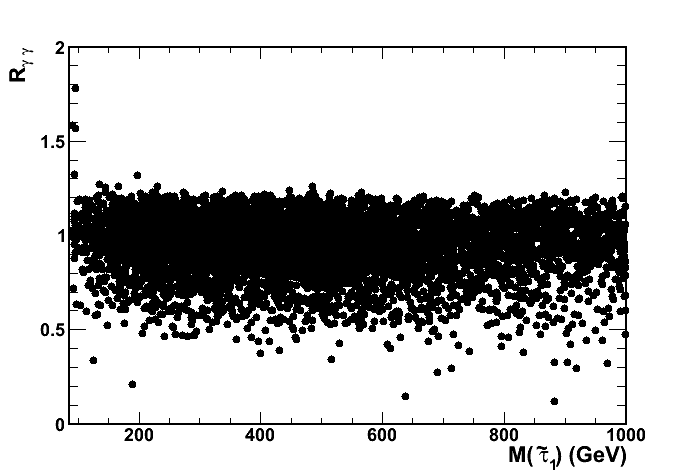} \\
\end{tabular}
\end{center}
\vspace*{-5mm}
\caption{\small  
$R_{\gamma\gamma}$ values for a sample of pMSSM points as a function of $m_{\chi^{\pm}_{1}}$ (left) 
and  ($m_{\tau^{\pm}_{1}}$) (right). We impose $R_{bb} >$ 0.9, to remove the effects due to the 
changes of the total width through the $bb$ channel.}
\label{fig:gamgam}
\end{figure}
The sbottom contribution to the $h\gamma\gamma$ vertex is  also very small,
for the same reasons discussed above in the case of the $hgg$ amplitude, and also 
because of its electric charge, $-\frac13$ compared to $+\frac23$ for stops. 
Other charged particles can also contribute to the $h\to \gamma \gamma$
rate \cite{ST-pp}. The charged Higgs bosons have negligible contributions for
$m_{H^\pm} \gsim 200$ GeV. Charginos contribute to the $h\gamma \gamma$ vertex and, 
because of their spin $\frac12$ nature, they contribution is only damped by powers 
of $M_h/m_{\chi^\pm}$. However, the $h\chi_{1,2}^\pm \chi_{1,2}^\mp$
couplings are similar in nature to those of the LSP given in eq.~(\ref{ghchi})
and cannot be strongly enhanced. As a result we expect contributions at most of the
order of 10\% even for mass values $m_{\chi_1^\pm} \approx 100$ GeV (see Fig.~\ref{fig:gamgam}).  
Charged sleptons have in general also little effect on  the $h\gamma\gamma$ vertex, with the 
exception of  staus~\cite{Carena:2011aa}. These behave like 
the bottom squarks. At very large $\mu\tb$ values, the splitting between the two $\tilde
\tau$ states  becomes significant and their couplings to the $h$ boson  large.
Since $\tilde \tau_1$ can have a mass of the order of a few 100~GeV, without
affecting the value of $M_h$, its contribution to the $h\gamma\gamma$ amplitude may 
be significant for largte values of $X_\tau$ (see Fig.~\ref{fig:gamgam}).  

\subsubsection{Impact of the LHC data} 

Now, it is interesting to perform a first assessment of the compatibility of the LHC and
Tevatron data with the MSSM and analyse the region of parameter favoured by the
observed boson mass and rate pattern (see also~\cite{Arbey:2011aa,Benbrik:2012rm}). 
Despite the preliminary character of the results reported by the LHC collaborations and the 
limited statistical accuracy of these first results, the study is a template for future 
analyses. In this analysis, we computing the $\chi^2$ probability on the observable 
of Table~\ref{tab:input} for each accepted pMSSM points. For the $b \bar b$ and $\tau^+ \tau^-$ 
channels, in which no evidence has been obtained at the LHC, we add the channel contribution 
to the total $\chi^2$ only when their respective $\mu$ value exceeded 1.5 and the pMSSM point 
becomes increasingly less consistent to the limits reported by CMS.
In order to investigate the sensitivity  to the inputs, we also compare the results by  including 
or not the $b \bar b$, for which a tension exists between the CMS limit and Tevatron results, 
and the $\tau^+ \tau^-$ rate. Figure~\ref{fig:pdf2D} shows the region of the $[X_t, m_{\tilde t_1}]$,
$[X_b, m_{\tilde b_1}]$ and $[M_A$, $\tan \beta]$  parameter space where pMSSM points are compatible with
the input $h$ boson mass and observed yields. In particular, we observe an almost complete suppression 
for low values of the sbottom  mixing parameter $X_b$.
\begin{figure}[h!]
\vspace*{-3mm}
\begin{center}
\begin{tabular}{ccc}
\includegraphics[width=5.75cm]{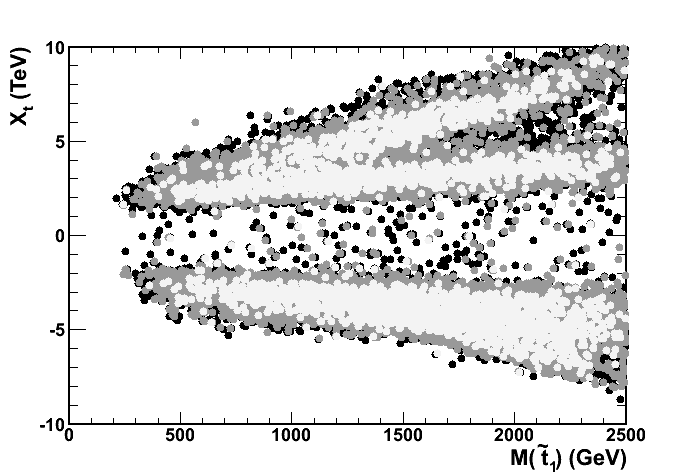} & 
\hspace*{-0.75cm} \includegraphics[width=5.75cm]{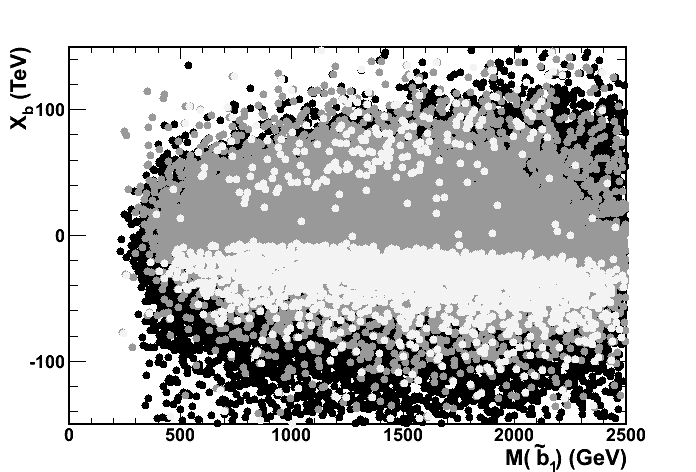} &
\hspace*{-0.75cm} \includegraphics[width=5.75cm]{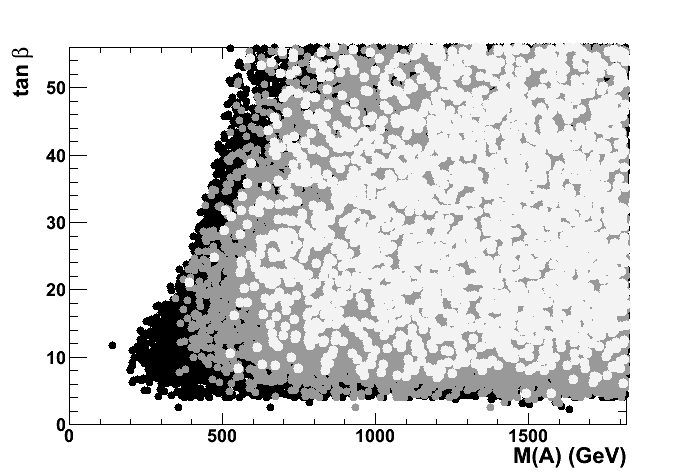} \\
\end{tabular}
\end{center}
\vspace*{-5mm}
\caption{\small Distributions of the pMSSM points in the $[X_t, m_{\tilde t_1}]$ 
(left), $[X_b, m_{\tilde b_1}]$ (centre) and $[M_A$, $\tan \beta]$ (right) parameter space. 
The black dots show the selected pMSSM points, those in light (dark) grey the same points 
compatible at 68\% (90\%) C.L. with the the Higgs constraints of Table~\ref{tab:input}.}
\label{fig:pdf2D}
\end{figure}

The distributions for some individual parameters which manifest a sensitivity are presented in Figure~\ref{fig:pdf1D}, where each pMSSM point 
enters with a weight equal to its $\chi^2$ probability. Points having a probability below 0.15 are not included. The probability weighted 
distributions obtained from this analysis are compared to the normalised frequency distribution for the same observables obtained for accepted 
points within the allowed mass region 122.5 $< M_H <$127.5~GeV. 
We observe that some variables are significantly affected by the constraints applied. Not surprisingly, the 
observable which exhibits the largest effect is the product $\mu \tan \beta$, for which the data favours large positive values, where 
the $\gamma \gamma$ branching fraction increases and the $b \bar b$ decreases as discussed above. On the contrary, it appears difficult 
to reconcile an enhancement of both $\mu_{\gamma \gamma}$ and $\mu_{b \bar b}$, as would be suggested by the central large value of 
$\mu_{b \bar b}$ = 1.97$\pm$0.72 recently reported by the Tevatron experiments~\cite{Tevatron:2012zz}. Such an enhancement is not observed 
by the CMS collaboration and the issue is awaiting the first significant evidence of a boson signal in the $b \bar b$ final state at the LHC 
and the subsequent rate determination. The $\tan \beta$ distribution is also shifted towards larger value as an effect of the Higgs mass and 
rate values. We also observe a significant suppression of pMSSM points with the pseudo-scalar $A$ boson mass below $\sim$450~GeV. 
This is due to the combined effect 
of the $A \rightarrow \tau^+ \tau^-$ direct searches and $B_s \rightarrow \mu^+ \mu^-$ rate, which constrain the [$M_A - \tan \beta$] plane 
to low $\tan \beta$ value for light $A$ masses, by the shift to $\mu \tan \beta$ from the Higgs rates disfavouring the low $\tan \beta$ region
and by the suppression of the non-decoupling regime.  

In quantitative terms, we observe that 0.06 (0.50) of the selected pMSSM points are compatible with the constraints given in 
Table~\ref{tab:input} at the 68\% (90\%) confidence level. If we remove the constraint on the upper limit constraint on the $b \bar b$ and 
$\tau^+ \tau^-$ rates, the fraction of points accepted at the 90\% C.L. does not change significantly, at 0.56, but that at the 68\% C.L. 
doubles to 0.12. On the contrary, if we replace the CMS upper limit for $\mu_{bb}$ with the $\mu_{bb}$ result of the Tevatron experiments 
for $M_H$ = 125~GeV~\cite{Tevatron:2012zz}, the fraction of accepted points at 68\% C.L. drops below 0.005. This highlights the tension which 
will be created in the pMSSM by a simultaneous excess in the $\gamma \gamma$ and $b \bar b$ channels, excess which cannot be adequately described 
in the pMSSM, as discussed above (see Figure~\ref{fig:bbgg}). 
\begin{figure}[h!]
\begin{center}
\begin{tabular}{cc}
\includegraphics[width=6.5cm]{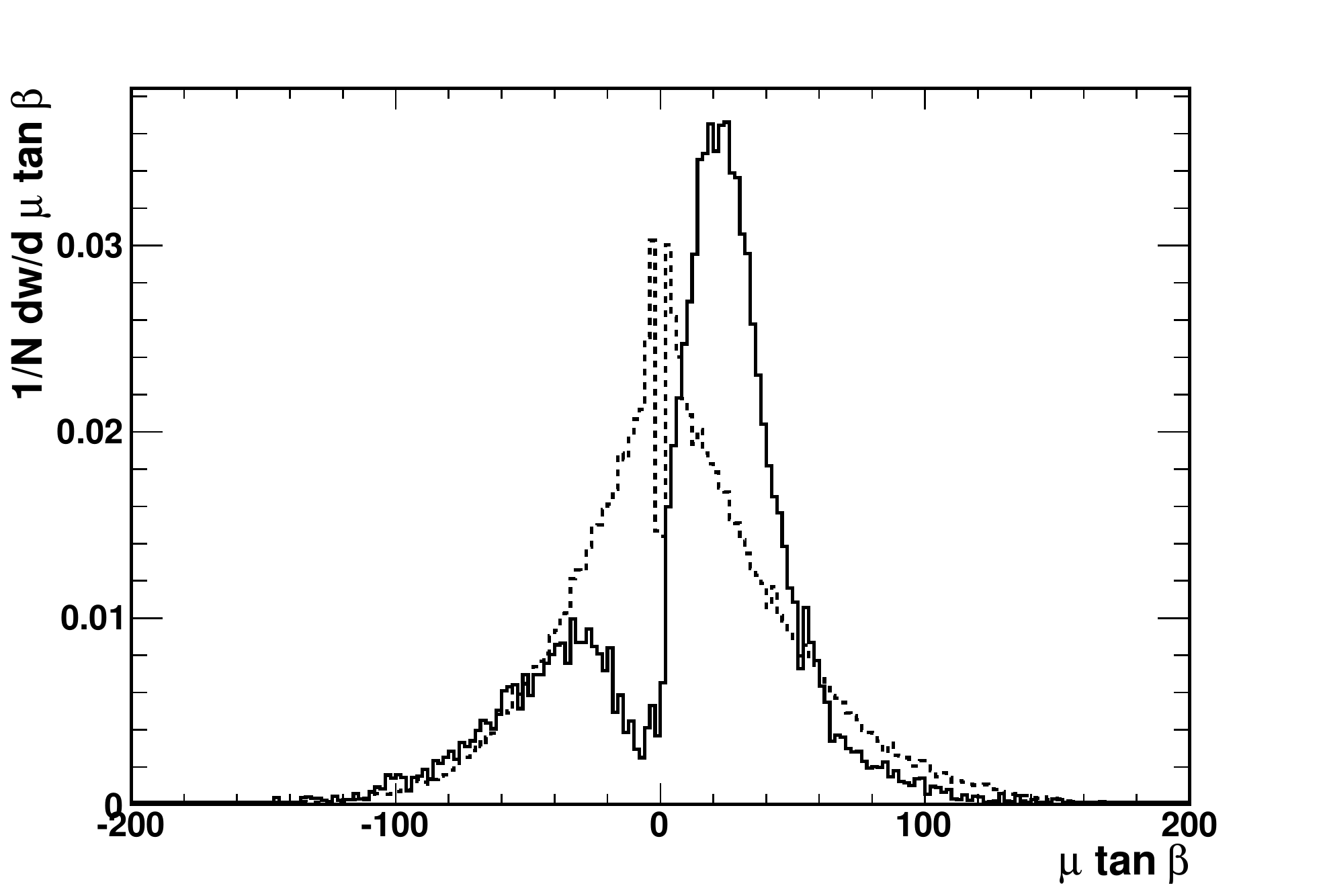} &
\includegraphics[width=6.5cm]{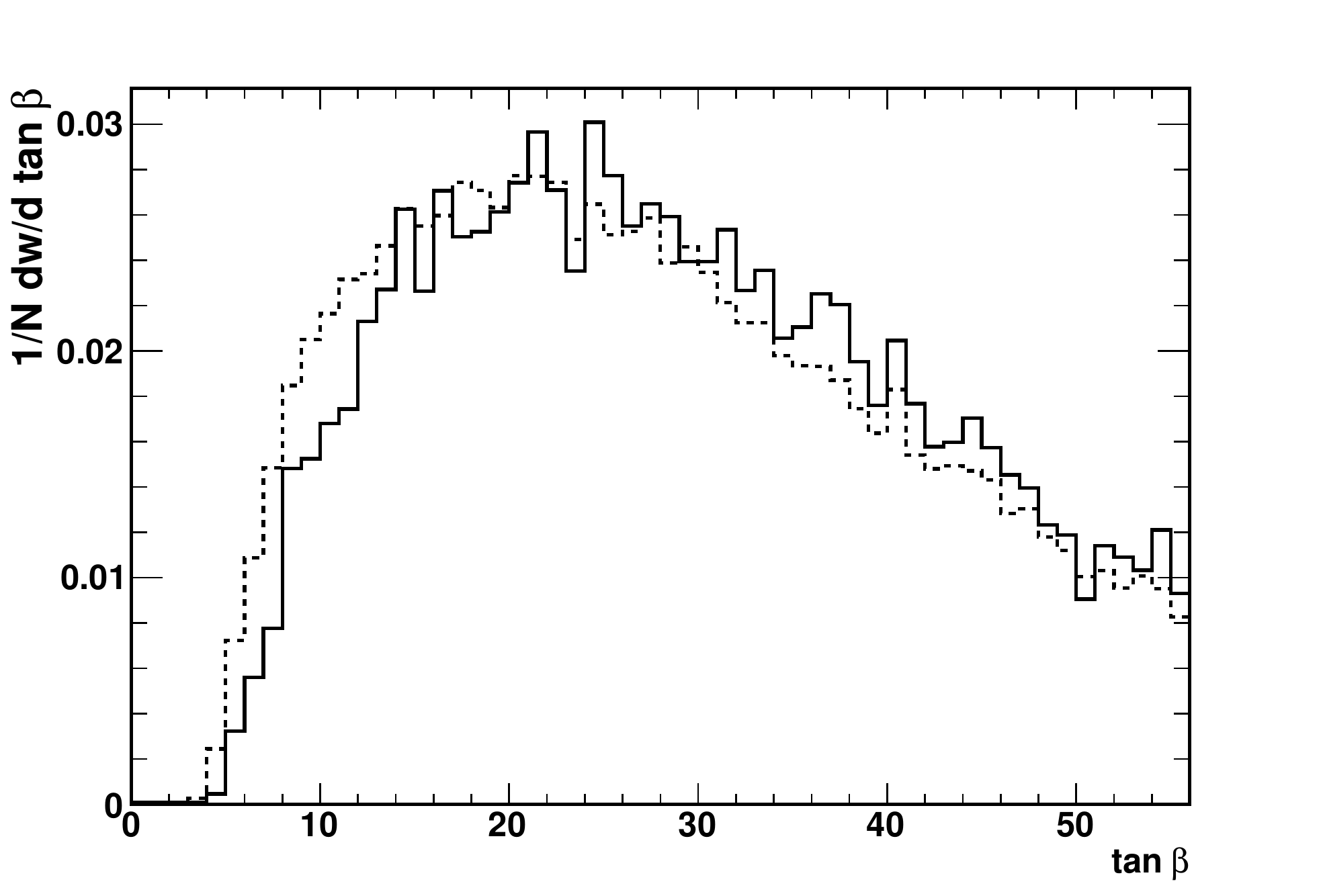} \\
\includegraphics[width=6.5cm]{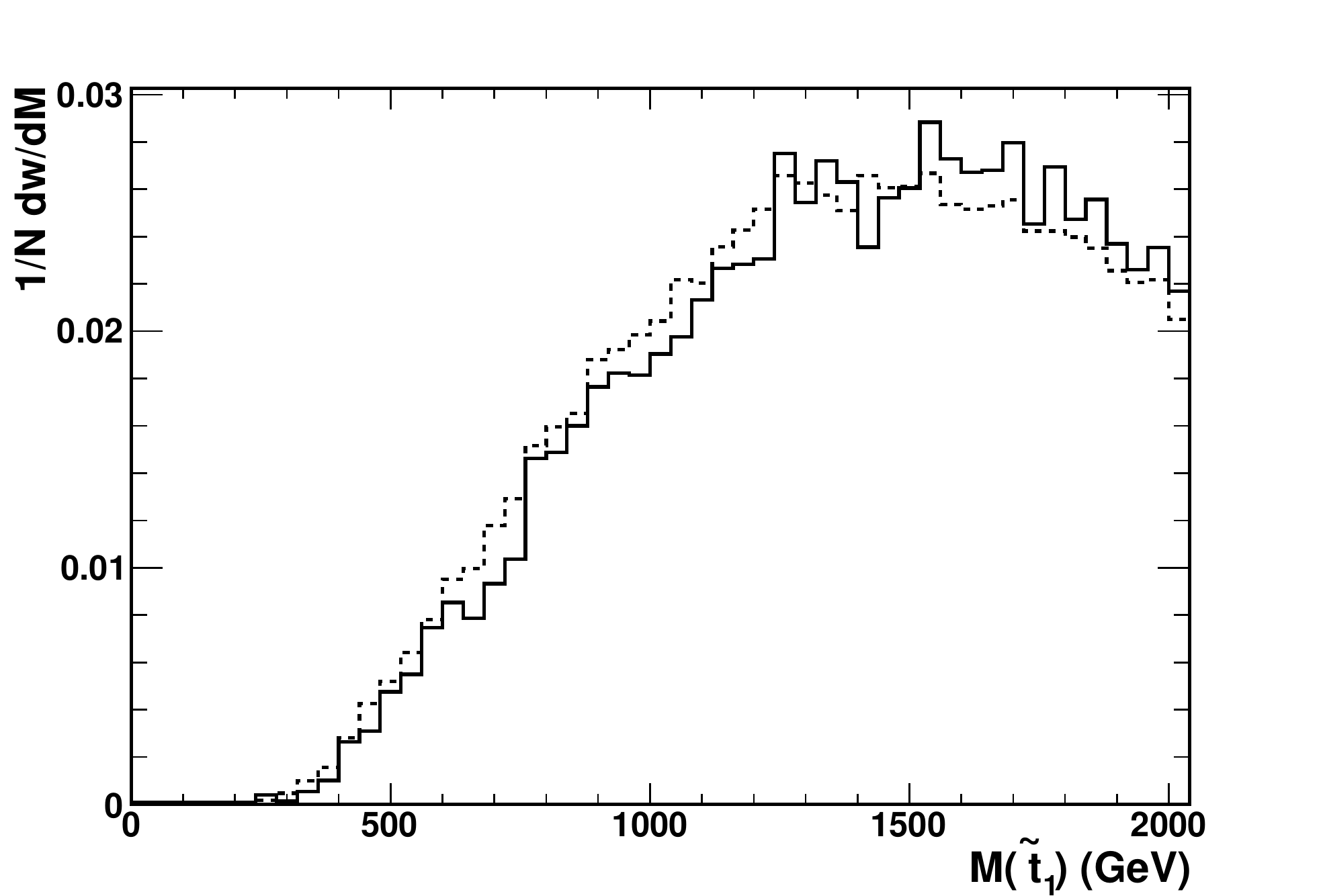} &
\includegraphics[width=6.5cm]{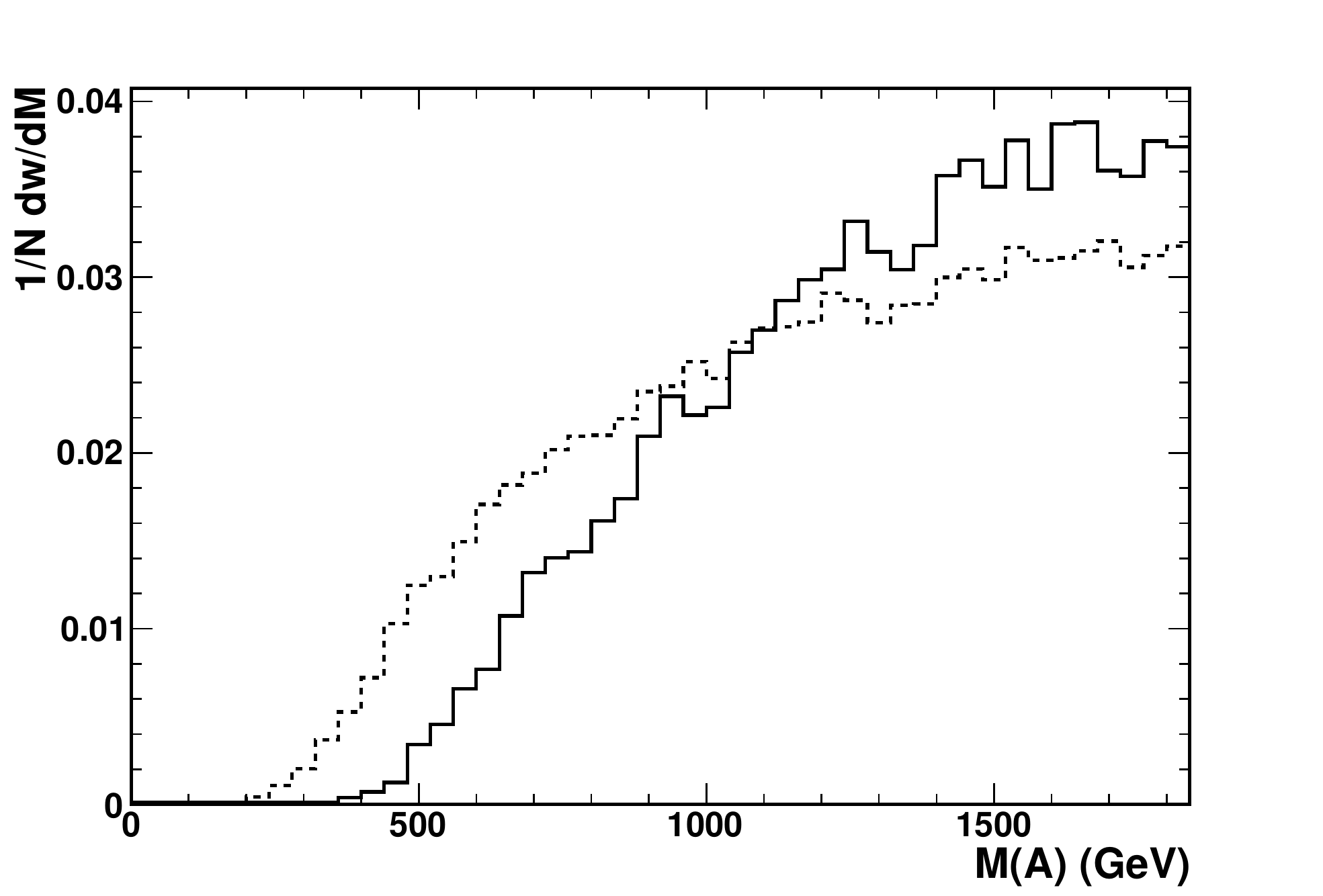} \\
\end{tabular}
\end{center}
\caption{\small The normalised distribution of the values of the $\mu \tan \beta$ 
(upper left), $\tan \beta$ (upper right), $M_{\tilde t_1}$ (lower left) and $M_{A}$ (lower right) variables for the 
selected pMSSM points (dashed line) compared to the probability density function for the same variables obtained from the  
$\chi^2$ probability using $M_h$, $R_{\gamma \gamma}$ and $R_{ZZ}$ (continuous line). The normalised distributions reflect 
the biases induced by the Higgs constraints.}
\label{fig:pdf1D}
\end{figure}

\section{Conclusions} 

The  implications of the new boson observation by the ATLAS and CMS collaborations for the phenomenological MSSM
have been outlined. The study has been based on broad scans over the pMSSM parameter space where points have been 
preselected based on constraints from electro-weak and flavour physics, dark matter and searches at LEP2 and the 
LHC. Various scenarios for the stop mixing parameter $X_t$ (maximal, typical and zero--mixing) and representative values 
of the soft SUSY--breaking scale  $M_S$ (1, 2 and 3 TeV) have been confronted with the Higgs mass range compatible with 
LHC results, accounting for systematic uncertainties. In order to obtain $M_h$ in the mass range 123 GeV$\leq M_h \leq$ 
129 GeV, large values of $M_S$ and/or $X_t$  are required. In particular, the $M_h$ constraints are sensitive to the 
value of the top quark mass for which the value extracted from the top quark pair production cross section has a more 
unambiguous definition but larger uncertainties. 

The various regimes of the pMSSM Higgs sector have been 
examined in the $[M_A, \tb]$ parameter. Of these regimes, only the decoupling regime, where the lighter $h$ boson 
has almost SM--like properties and the heavier Higgs particles decouple from gauge bosons, and the SUSY regime survives all
constraints. The anti-decoupling regime where the $H$ state plays the role of the SM Higgs boson,  the  intense coupling 
regime in which there are three light states $h,H$ and $A$, the vanishing coupling regime in which the $h$ coupling to
bottom quarks or gauge bosons are very strongly suppressed,  and most of the intermediate coupling regime with relatively 
low $M_A$ and $\tb$ values, are excluded by the present data. In the SUSY regime light superparticles may affect 
the production and decay rates of the $h$ boson. Light neutralinos may lead to invisible $h$ boson decays, light stop 
and sbottom quarks affect the $h b\bar b$ couplings and the production cross  section in the dominant gluon-gluon 
fusion mechanism, and light squarks, $\tau$-sleptons and charginos  may affect the $h \to \gamma \gamma$ decay mode. 

We have confronted these possibilities with the recent LHC results and find that a significant fraction of pMSSM points 
in our scan compatible with them, including a possible enhancement of the $\gamma \gamma$ rate.
Improved  precision in the experimental measurements and sensitivity to the direct searches for the heavier Higgs bosons 
and supersymmetric particle partners at the LHC will provide the basis for clarifying the relation between the newly 
discovered scalar sector and physics beyond the Standard Model. 

\subsubsection*{Acknowledgements}

A.A. and F.M. acknowledge partial support from the European Union FP7
ITN INVISIBLES (Marie Curie Actions, PITN-GA-2011-289442).
A.D. thanks the CERN TH unit for hospitality.


\begin{thebibliography}{99}

\bibitem{ATLAS:2012zz}
  [ATLAS Collaboration], CERN-PH-EP-2012-167 and
  F.~Gianotti, for the ATLAS Collaboration,
  CERN Seminar, July 4th, 2012.
  
\bibitem{CMS:2012zz}
  [CMS Collaboration], CMS PAS HIG-12-020 and
  J.~Incandela, for the CMS Collaboration,
  CERN Seminar, July 4th, 2012.

\bibitem{Tevatron:2012zz}
  The CDF and D0 Collaborations,
  FERMILAB-CONF-12-318-E.

\bibitem{ABDMQ}
  A.~Arbey, M.~Battaglia, A.~Djouadi, F.~Mahmoudi and J.~Quevillon,
  Phys.\ Lett.\ B708 (2012) 162. 

\bibitem{Arbey:2011aa}
  A.~Arbey, M.~Battaglia and F.~Mahmoudi,
  Eur.\ Phys.\ J.\ C72 (2012) 1906.

\bibitem{Carena:2011aa}
  M.~Carena, S.~Gori, N.~R.~Shah and C.~E.~M.~Wagner,
  JHEP 1203 (2012) 014.

\bibitem{papers-2011}
  H.~Baer, V.~Barger and A.~Mustafayev,
  Phys.\ Rev.\ D85 (2012) 075010;
%
  S.~Heinemeyer, O.~Stal and G.~Weiglein,
  Phys.\ Lett.\ B710 (2012) 201;
%
  P.~Draper, P.~Meade, M.~Reece and D.~Shih,
  Phys.\ Rev.\ D85 (2012) 095007;
%
%
  O.~Buchmueller, R.~Cavanaugh, A.~De Roeck, M.~J.~Dolan, J.~R.~Ellis, H.~Flacher, S.~Heinemeyer and G.~Isidori {\it et al.},
  arXiv:1112.3564 [hep-ph];
%
  S.~Akula, B.~Altunkaynak, D.~Feldman, P.~Nath and G.~Peim,
  Phys.\ Rev.\ D85 (2012) 075001
  C.~Strege, G.~Bertone, D.~G.~Cerdeno, M.~Fornasa, R.~R.~de Austri and R.~Trotta,
  JCAP 1203 (2012) 030;
  C.~Beskidt, W.~de Boer, D.~I.~Kazakov and F.~Ratnikov,
  JHEP 1205 (2012) 094;
%
  M.~Carena, S.~Gori, N.~R.~Shah, C.~E.~M.~Wagner and L.~-T.~Wang,
  arXiv:1205.5842 [hep-ph];
  M.~W.~Cahill-Rowley, J.~L.~Hewett, A.~Ismail and T.~G.~Rizzo,
  arXiv:1206.5800 [hep-ph];

\bibitem{pMSSM} A. Djouadi and S. Rosiers--Lees (conv.) et al., Summary  Report
  of the MSSM Working Group for the ``GDR--Supersym\'etrie",  hep-ph/9901246.

\bibitem{Review2} A. Djouadi, Phys. Rept. 459 (2008) 1.

\bibitem{rad-cor}  M.\, Carena and H. 
Haber, Prog.\ Part.\ Nucl.\ Phys. 50 (2003) 63;  S. Heinemeyer, W. Hollik and G.
Weiglein, Phys. Rept. 425 (2006) 265; S. Heinemeyer, Int. J. Mod. Phys A21
(2006) 2659. 

\bibitem{Review1}  A.~Djouadi, Phys. Rept.  457 (2008) 1.

\bibitem{RC-1loop} Y.~Okada, M.~Yamaguchi and T.~Yanagida, Prog.\ Theor.\
Phys.\  85 (1991) 1;  J.R.~Ellis, G.~Ridolfi and F.~Zwirner, Phys.\ Lett.\ B257
(1991) 83;  H.E.~Haber and R.~Hempfling, Phys.\ Rev.\ Lett.\ 66  (1991) 1815.

\bibitem{benchmarks} M. Carena, S. Heinemeyer, C. Wagner and G. Weiglein, Eur.
Phys. J. C26 (2003) 601. 

 \bibitem{adkps} B. Allanach et al., JHEP 0409 (2004) 044.  

\bibitem{BDSZ} G.~Degrassi, P.~Slavich and F.~Zwirner, Nucl.\ Phys.\ B611 (2001)
403; A.~Brignole, G.~Degrassi, P.~Slavich and F.~Zwirner, Nucl.\ Phys.\ B643
(2002) 79; {\it ibid.}  B631 (2002) 195;

\bibitem{suspect} A. Djouadi, J.L. Kneur and G. Moultaka,  Comput. Phys. Commun.
176 (2007)  426.

\bibitem{Allanach:2001kg}
  B.~C.~Allanach,
  Comput.\ Phys.\ Commun.\  143 (2002) 305.

\bibitem{HHW} See for instance, S.~Heinemeyer, W.~Hollik and G.~Weiglein,
Phys. Rev. D58 (1998) 091701  and Eur.\ Phys.\ J.C9 (1999) 343. 

\bibitem{Heinemeyer:1998yj}
  S.~Heinemeyer, W.~Hollik and G.~Weiglein,
  Comput.\ Phys.\ Commun.\  124 (2000) 76.

\bibitem{3loop} 
  P. Kant, R. Harlander, L. Mihaila and M. Steinhauser, 
  JHEP 1008 (2010) 104.

\bibitem{Decoupling} 
  H.E. Haber and Y. Nir, Phys. Lett. B306 (1993) 327; 
  H.E. Haber, hep-ph/9505240.  

\bibitem{Antidecoup} 
  J.F. Gunion, A. Stange, S. Willenbrock 
  et al., hep-ph/9602238. 

\bibitem{intense}  
  E. Boos et al., Phys. Rev. D66 (2002) 055004; E. Boos et
  al., Phys. Lett. B622 (2005) 311; A. Djouadi and Y. Mambrini, JHEP 0612
  (2006) 001; E. Boos, A. Djouadi and A. Nikitenko, Phys. Lett. B578  (2004)
  384. 

\bibitem{vanishing} 
  H. Baer and J. Wells, Phys. Rev. D57 (1998) 4446; W. Loinaz
  and J.D. Wells, Phys. Lett. B445 (1998) 178; K.S. Babu and C.F. Kolda, Phys.
  Lett. B451 (1999) 77;  M.~Carena, S.~Mrenna and C.E.M.~Wagner, Phys. Rev. D62 
  (2000) 055008; S.~Heinemeyer, W.~Hollik and G.~Weiglein, Eur. Phys. J. C16 
  (2000) 139; D. Noth and M. Spira,   JHEP 1106 (2011) 084.

\bibitem{ST-gg} A.~Djouadi, Phys. Lett. B435 (1998) 101. 

\bibitem{ST-pp} A. Djouadi, V.~Driesen, W.~Hollik and J.~I.~Illana, Eur. Phys.
J.C1 (1998) 149.

\bibitem{villadoro}
  A.~Arvanitaki and G.~Villadoro,
  JHEP 1202 (2012) 144.

\bibitem{H-LSP} 
  H. K. Dreiner, J.S. Kim and
  O. Lebedev, arXiv:1206.3096. See also,  A.~Djouadi et al.,  Phys. Lett. B376
  (1996) 220;  Z.~Phys.~C70 (1996) 435; Z. Phys. C74 (1997) 93; A. Djouadi, Mod.
  Phys. Lett. A14 (1999) 359.

\bibitem{Djouadi:1997yw}
  A.~Djouadi, J.~Kalinowski and M.~Spira,
  Comput.\ Phys.\ Commun.\  {108} (1998) 56.
  
\bibitem{AlbornozVasquez:2011aa}
  D.~Albornoz Vasquez, G.~Belanger, R.~Godbole and A.~Pukhov,
  arXiv:1112.2200 [hep-ph].
 
\bibitem{bb-nnlo} 
  R. Harlander and W. Kilgore, Phys.  Rev. D68 (2003) 013001.

\bibitem{PDG} Particle Data Group (K. Nakamura et al.),  J. Phys. G37 (2010)
075021.

\bibitem{Arbey:2011un}
  A.~Arbey, M.~Battaglia and F.~Mahmoudi,
  Eur.\ Phys.\ J.\ C72 (2012) 1847.

\bibitem{sdecay} M. Muhlleitner, A. Djouadi and Y. Mambrini, Compt. Phys.
  Com. 168 (2005) 46. It is based on:   C. Boehm, A. Djouadi and Y.
  Mambrini,  Phys.Rev. D61 (2000) 095006;  A. Djouadi and Y. Mambrini, Phys.
  Rev. D63 (2001) 115005; A.~Djouadi, Y.~Mambrini and M.~Muhlleitner, Eur. 
  Phys. J. C20 (2001) 563.

\bibitem{flavor}
  F.~Mahmoudi,
  Comput.\ Phys.\ Commun.\ 178 (2008) 745;
  F.~Mahmoudi,
  Comput.\ Phys.\ Commun.\  180 (2009) 1579;
  A.~Arbey and F.~Mahmoudi,
  Comput.\ Phys.\ Commun.\  181 (2010) 1277.

\bibitem{higlu}
  M.~Spira, Nucl.\ Instrum.\ Meth.\ A389 (1997) 357; hep-ph/9510347.
  Based on, M.~Spira, A.~Djouadi, D.~Graudenz and P.~M.~Zerwas,
  Nucl.\ Phys.\ B453 (1995) 17.


\bibitem{Aaltonen:2011fi}
  T.~Aaltonen {\it et al.}  [CDF Collaboration],
  Phys.\ Rev.\ Lett.\  107 (2011) 239903;
   [Phys.\ Rev.\ Lett.\  107 (2011) 191801].

\bibitem{Aaij:2012ac}
  R.~Aaij {\it et al.}  [LHCb Collaboration],
  Phys.\  Rev.\  Lett.\  108 (2012) 231801.

\bibitem{Chatrchyan:2012rg}
  S.~Chatrchyan {\it et al.}  [CMS Collaboration],
  JHEP 1204 (2012) 033.

\bibitem{Akeroyd:2011kd}
  A.~G.~Akeroyd, F.~Mahmoudi and D.~M.~Santos,
  JHEP 1112 (2011) 088;
%
  F.~Mahmoudi, S.~Neshatpour and J.~Orloff,
  arXiv:1205.1845 [hep-ph].

\bibitem{Hurth:2012jn}
   T.~Hurth and F.~Mahmoudi,
   arXiv:1207.0688 [hep-ph].

\bibitem{Aprile:2011hi}
  E.~Aprile {\it et al.}  [XENON100 Collaboration],
  Phys.\ Rev.\ Lett.\  107 (2011) 131302.

\bibitem{Komatsu:2010fb}
  E.~Komatsu {\it et al.}  [WMAP Collaboration],
  Astrophys.\ J.\ Suppl.\  192 (2011) 18.

\bibitem{Arbey:2008kv}
  A.~Arbey and F.~Mahmoudi,
  Phys.\ Lett.\ B669 (2008) 46;
  A.~Arbey and F.~Mahmoudi,
  JHEP 1005 (2010) 051.

\bibitem{Aad:2011ib}
  G.~Aad {\it et al.}  [ATLAS Collaboration],
  Phys.\ Lett.\ B710 (2012) 67.

\bibitem{Chatrchyan:2011zy}
  S.~Chatrchyan {\it et al.}  [CMS Collaboration],
  Phys.\ Rev.\ Lett.\  107 (2011) 221804.

\bibitem{Aad:2011rv}
  G.~Aad {\it et al.}  [ATLAS Collaboration],
  Phys.\ Lett.\ B705 (2011) 174.

\bibitem{Chatrchyan:2012vp}
  S.~Chatrchyan {\it et al.}  [CMS Collaboration],
  Phys.\ Lett.\ B713 (2012) 68.

\bibitem{Aad:2011hh}
  G.~Aad {\it et al.}  [ATLAS Collaboration], JHEP 06 (2012) 039.

\bibitem{ATLAS-2012-091}
  [ATLAS Collaboration], Note ATLAS-CONF-2012-091.

\bibitem{CMS-12-015}
  [CMS Collaboration], Note CMS PAS HIG-2012-015.

\bibitem{ATLAS-2012-092}
  [ATLAS Collaboration], Note ATLAS-CONF-2012-092.

\bibitem{CMS-12-016}
  [CMS Collaboration], Note CMS PAS HIG-2012-016.

\bibitem{CMS-12-019}
  [CMS Collaboration], Note CMS PAS HIG-2012-019.

\bibitem{CMS-12-018}
  [CMS Collaboration], Note CMS PAS HIG-2012-018.

\bibitem{THU} S. Dittmaier {\it et al.}, [LHC Higgs cross section working group], 
  arXiv:1101.0593 [hep-ph].

\bibitem{JB}  J. Baglio and A. Djouadi, JHEP 1103 (2011) 055. 

\bibitem{Lancaster:2011wr}
  [Tevatron Electroweak Working Group and CDF and D0 Collaborations],
  arXiv:1107.5255 [hep-ex].

\bibitem{pole-mt} S. Alekhin, A. Djouadi and S. Moch, arXiv:1207.0980 [hep-ph].  

\bibitem{Hpole}  H. Baer, A. Belyaev, T. Krupovnickas  and X. Tata, JHEP 0402
(2004) 007; JHEP 0406 (2004) 061;  
  A.~Djouadi, M.~Drees and J.-L.~Kneur,
  Phys.\ Lett.\ B624 (2005) 60;
  [hep-ph/0504090].

\bibitem{Arbey:2012na}
  A.~Arbey, M.~Battaglia and F.~Mahmoudi,
  arXiv:1205.2557 [hep-ph].

\bibitem{Benbrik:2012rm}
  R.~Benbrik, M.~G.~Bock, S.~Heinemeyer, O.~Stal, G.~Weiglein and L.~Zeune,
  arXiv:1207.1096 [hep-ph].

\end{thebibliography}
\end{document}